\documentclass[12pt, draftclsnofoot, onecolumn]{IEEEtran}
\usepackage{amsmath, graphics,amssymb,epsfig,subfigure}
\usepackage{etoolbox}
\usepackage{cite,xcolor}
\usepackage{url}
\usepackage{algorithm,algorithmic}
\usepackage{color,soul}
\usepackage{cases}
\usepackage{graphicx,amsmath}
\usepackage{stfloats}
\usepackage{epstopdf}
\usepackage{multirow}
\usepackage{comment}
\usepackage{tcolorbox}
\makeatletter
\patchcmd{\@makecaption}
  {\scshape}
  {}
  {}
  {}
\makeatother
\newtheorem{assumption}{Assumption}

\newtheorem{lemma}{Lemma}
\newtheorem{theorem}{Theorem}

\newtheorem{definition}{Definition}

\begin{document}
\bstctlcite{IEEEexample:BSTcontrol}
\title{On the Convergence of Large Language Model Optimizer for Black-Box Network Management}
\author{Hoon Lee,~\IEEEmembership{Member,~IEEE}, Wentao Zhou,\\Merouane Debbah,~\IEEEmembership{Fellow,~IEEE}, and Inkyu Lee,~\IEEEmembership{Fellow,~IEEE}
\vspace{-5mm}
\thanks{

H. Lee is with the Department of Electrical Engineering and the Artificial Intelligence Graduate School, Ulsan National Institute of Science and Technology (UNIST), Ulsan, 44919, South Korea. 

W. Zhou and I. Lee are with the School of Electrical Engineering, Korea University, Seoul 02841, Korea (e-mail: inkyu@korea.ac.kr). 

M. Debbah is with Khalifa University of Science and Technology, PO Box 127788, Abu Dhabi, UAE.
}}

\maketitle
\begin{abstract}
Future wireless networks are expected to incorporate diverse services that often lack general mathematical models. To address such black-box network management tasks, the large language model (LLM) optimizer framework, which leverages pretrained LLMs as optimization agents, has recently been promoted as a promising solution. 
This framework utilizes natural language prompts describing the given optimization problems along with past solutions generated by LLMs themselves. 
As a result, LLMs can obtain efficient solutions autonomously without knowing the mathematical models of the objective functions. Although the viability of the LLM optimizer (LLMO) framework has been studied in various black-box scenarios, it has so far been limited to numerical simulations. For the first time, this paper establishes a theoretical foundation for the LLMO framework. With careful investigations of LLM inference steps, we can interpret the LLMO procedure as a finite-state Markov chain, and prove the convergence of the framework. 
Our results are extended to a more advanced multiple LLM architecture, where the impact of multiple LLMs is rigorously verified in terms of the convergence rate. 
Comprehensive numerical simulations validate our theoretical results and provide a deeper understanding of the underlying mechanisms of the LLMO framework.
\end{abstract}

\section{Introduction}
Future wireless communication systems have become increasingly complex and specific, requiring diverse performance indicators. To meet unique demands arising from heterogeneous system requirements, it is essential to design proper network optimization algorithms for a wide range of applications. Normally, this necessitates mathematical models of network objective functions, particularly closed-form expressions that define the network performance. However, acquiring accurate models for various service applications is infeasible. Furthermore, practical wireless communication networks encompass several issues, such as imperfect channel knowledge \cite{NLee:20,WZhou:24} and hardware imperfections \cite{BEmil:15}, for which mathematical models are generally unavailable.

As a result, performance indicators of future wireless systems often lack closed-form expressions, which leads to black-box optimization (BBO) tasks \cite{BBO}. Existing network optimization approaches, including convex/nonconvex optimization algorithms \cite{Boyd,BSUM} and deep learning-based solutions \cite{HLee:19JSAC,HLee:22,HLee:22-IoT,HLee:21,HLee:21-TWC}, heavily rely on analytical formulas for gradients and Hessians, and thus they cannot address the problems whose objective functions are not given in closed-form. 

Such problems may be solved by traditional BBO techniques including genetic algorithms (GA) \cite{GA}, Bayesian optimization (BO) \cite{BO}, and model-free reinforcement learning (RL) methods \cite{RL}. 
These approaches utilize the objective function values obtained from numerical simulations or real-world measurements, thereby eliminating the need for explicit mathematical expressions. They depend on optimization agents that are implemented through stochastic search mechanisms \cite{GA,BO}, machine learning models \cite{BO}, and neural networks (NNs) for deep RL methods. 
Traditional BBO methods require extensive human intervention for determining hyperparameters, such as learning rate, optimization algorithms, NN architectures, and training procedures. Since the performance of BBO techniques highly depends on these man-made components, it is necessary to fine-tune hyperparameters carefully. In general, there are no systematic ways for identifying optimal hyperparameters, resulting in trial-and-error-based search procedures. Since optimal choices of hyperparameters vary depending on the objective/constraint functions and network inputs, their optimization procedures should be tailored to each network configuration. Consequently, these methods lack the generalization ability of a universal network solver.

Recently, there has been innovative research based on the large language model (LLM) optimizer framework, which utilizes LLMs, i.e., generative pretrained transformer (GPT) \cite{GPT2,GPT3} and LLM Meta AI (LLaMA) \cite{LLaMA}, as BBO agents. The effectiveness of such an LLM optimizer (LLMO) framework has been demonstrated in various optimization domains \cite{OPRO,BHuang:24,LMEA,GD,ANie:23,LEO,MOEAD,HLee:24a,HLee:24b,MALLM,MALLM2,HLi:24,KQiu:24,JTong:24}. 
LLMs have been proven to be effective in addressing various types of sequential data processing, including natural language processing \cite{GPT3,CoT,CoT-SC}, signal estimation \cite{mirchandani2023large}, and sensing applications \cite{LLMSense}. Recent works have revealed that LLMs can fairly improve the quality of responses by observing feedback on their past outputs \cite{ReACT,Reflextion,OPTIMUS}. Such a property enables intelligent workflows that control the behaviors of LLMs through feedback. Trained on massive and diverse corpora, pretrained LLMs can develop an internal policy using several examples given in input prompts. This in-context learning (ICL) ability allows them to handle a wide variety of wireless resource allocation and management tasks without retraining \cite{GPT3,CoT,CoT-SC}. These results promote LLMs as viable BBO agents in network optimization problems.

In the LLMO, rewards for past LLM outputs are provided in new prompts and LLMs are instructed to generate enhanced solutions.
Iteratively prompting LLMs with historical decisions allows them to learn effective optimization policies. Compared to existing BBO techniques that heavily resort to careful hyperparameter tuning, no human intervention is required for the LLMO since it relies only on internal reasoning capabilities of pretrained LLMs. Therefore, the LLMO has been regarded as a promising solution that implements a versatile network solver in a fully automated manner.

\subsection{Related works}
The LLMO technique was first introduced in \cite{OPRO}, where its capabilities for identifying the globally optimal solutions were demonstrated in simple convex optimization problems. Use cases in more complicated optimization tasks were studied in \cite{BHuang:24}. The LLMO exhibits promising performance in several cases and outperforms traditional BBO algorithms. However, it has been revealed that the LLMO generally struggles in unexplored solution spaces, making it challenging to address generic nonconvex optimization tasks.

This difficulty can be resolved by providing additional information in prompts. LLMs are designed to execute operations of well-known optimization algorithms, such as the GA \cite{LMEA} and gradient descent methods \cite{GD}. In addition, the work in \cite{ANie:23} incorporated gradient vectors of objective functions directly in input prompts. Notably, these methods perform better than conventional LLMO \cite{OPRO}. However, such approaches require human intervention in crafting input prompts \cite{LMEA,GD} and precise models of the objective functions \cite{ANie:23}. 

The full potential of LLMs has been explored in \cite{LEO,MOEAD}. The authors in \cite{LEO} introduced a new LLMO approach that utilizes two distinct LLMs interacting with each other. Inspired by \cite{BHuang:24}, this method seeks to overcome the limited exploration capabilities of the LLMO framework. Each LLM is assigned to either exploit or explore candidate solutions. Consequently, the autonomy of LLMs can be efficiently leveraged to tackle any given nonconvex optimization problem.

There have been recent studies on the LLMO for various BBO tasks in wireless communication networks, including resource allocation \cite{HLee:24a,HLee:24b,MALLM,MALLM2}, UAV trajectory optimization \cite{HLi:24}, access point placement \cite{KQiu:24}, and network slicing \cite{JTong:24}. The two-LLM architecture presented in \cite{LEO} has been extended to a generic multi-LLM structure \cite{HLee:24a}. This multi-LLMO approach for handling nonconvex resource management problems has significantly improved compared to single LLM counterparts \cite{OPRO}. It has been reported from \cite{HLee:24a} that the LLMO can achieve almost identical performance to existing nonconvex optimization algorithms that rely on mathematical models.

The work in \cite{KQiu:24} considered a network coverage map as an objective function that can only be evaluated through simulators. Also, \cite{JTong:24} addressed network slicing problems, which aim to assign individual users to appropriate service slices based on their quality-of-service requirements. 
In these cases, due to the absence of mathematical models, evaluating the objective values resorts to computationally expensive simulations. The LLMO has been reported to be powerful for handling such complicated network management tasks. 

Recent studies have combined the LLMO with traditional BBO algorithms, e.g., the GA \cite{LMEA, LEO, MOEAD, HLi:24, KQiu:24} and multi-agent reinforcement learning architecture \cite{HLee:24a,HLee:24b,MALLM,MALLM2}. The core idea of these works is to design efficient LLMO mechanisms inspired by classical BBO techniques. By doing so, noticeable performance gains can be achieved compared to the original approach \cite{OPRO}. However, they have been limited to exploring new use cases in wireless communication networks and developing heuristics for the LLMO. There have been no fundamental studies that rigorously analyze the performance of the LLMO in generic black-box network management tasks. It still remains unaddressed which features and conditions of the LLMO contribute to BBO network management tasks. 

\subsection{Contributions and organization}

For the first time, this paper presents the theoretical foundation of the LLMO by analyzing its optimality and convergence properties. Unlike the conventional works \cite{OPRO,BHuang:24,LMEA,GD,ANie:23,LEO,MOEAD,HLee:24a,HLee:24b,MALLM,MALLM2,HLi:24,KQiu:24,JTong:24} which simply design heuristics of the LLMO, this paper establishes the optimality of the LLMO for generic nonconvex BBO tasks in wireless networks. Our focus is particularly on establishing the theoretical analysis of the original LLMO architecture \cite{OPRO} and its variant \cite{HLee:24a}. The main contributions are summarized as follows:
\begin{itemize}
    \item With careful investigation of LLMs, we demonstrate that LLMs treat optimization variables as discrete representations within a finite language space. Consequently, the LLMO can be modeled as a Markov chain where new solutions are identified based on past decisions.
    \item Our analysis reveals that decision-making procedures of the LLMO can be characterized by a finite-state Markov chain model. Accordingly, the LLM behaves as a stochastic agent that maps past solutions in the input to new candidate solutions. This provides theoretical guarantees on the convergence, that is, solutions generated by LLMs are shown to converge to the global optimum. Based on this analysis, we can derive the necessary conditions for convergence, which provide guidelines for designing an efficient LLMO.
    \item Our analysis is extended to a generic multi-LLM scenario \cite{LEO,HLee:24a}, where several LLMs collaboratively solve BBO problems by exchanging their decision history. We rigorously prove the optimality and convergence behaviors of this multi-LLMO method and examine the impact of the number of LLMs on the convergence speed.
    \item We validate the theoretical results through extensive numerical simulations. The viability of the LLMO is verified in various networking problems, including resource management in interference channels (IFCs), multi-user broadcast channels (BCs), and massive multiple-input multiple-output (MIMO) systems. In these application scenarios, the LLMO is shown to generate almost identical performance to existing optimization algorithms. 
\end{itemize}

The rest of this paper is organized as follows: Section \ref{sec:sec2} details the operations of the LLMO framework. Convergence with a single LLM is analyzed in Section \ref{sec:sec3}. Section \ref{sec:sec4} presents an extension to a generic multi-LLM scenario. Section \ref{sec:sec5} demonstrates the LLMO through numerical simulations. Finally, Section \ref{sec:sec6} provides concluding remarks.

\textit{Notations:} We denote scalars, vectors, and matrices by lowercase normal symbols, lowercase boldface symbols, and uppercase boldface symbols, respectively. Sets of $U\times V$ real-valued matrices and column vectors of length $U$ are defined as $\mathbb{R}^{U\times V}$ and $\mathbb{R}^{U}$, respectively. Also, $\mathbf{0}_{U\times V}$ and $\mathbf{I}_{U}$ respectively indicate an all-zero matrix of size $U\times V$ and an identity matrix of size $U\times U$. All-zero and all-one column vectors of length $U$ are respectively denoted by $\mathbf{0}_{U}$ and $\mathbf{1}_{U}$. For a set $\mathbb{U}$, $|\mathbb{U}|$ stands for the cardinality. Table \ref{tab:tab1} summarizes acronyms and notations utilized in this paper.

\begin{table*}[t]
\centering
\caption{Acronyms and Notations}
\label{tab:tab1}
\begin{tabular}{|c|l||c|l|}
\hline
\textbf{Acronym} & \textbf{Description} & \textbf{Notations} & \textbf{Description} \\ \hline\hline
BBO   & Black‑box optimization & $\mathbf{x}$ & Action vector \\ \hline
BC    & Broadcast channel & $r(\cdot)$   & Reward function \\ \hline
BO    & Bayesian optimization & $\mathbf{X}^{(t)}$ & Action population at iteration $t$ \\ \hline
BS    & Base station & $\mathbf{r}^{(t)}$ & Reward vector at iteration $t$ \\ \hline
CoT   & Chain‑of‑thought & $[\mathbf{X}_{\sf{ex}}^{(t)},\mathbf{r}_{\sf{ex}}^{(t)}]$ & In-context example at iteration $t$ \\ \hline
NN   & Neural network & $\text{pmpt}^{(t)}$ & Prompt at iteration $t$ \\ \hline
EE    & Energy efficiency  &  $[\mathbf{x}^{(t)}_{\text{best}},r^{(t)}_{\text{best}}]$ & Best action-reward pair at iteration $t$ \\ \hline
GA    & Genetic algorithm & $\mathbb{M}^{(t)}$ & Memory at iteration $t$ \\ \hline
GNN   & Graph neural network & $\mathcal{P}(\cdot)$   & Prompt generator \\ \hline
ICL   & In‑context learning & $\mathcal{L}(\cdot)$   & LLM inference \\ \hline
IFC   & Interference channel & $\mathcal{S}(\cdot)$   & Sampling operator \\ \hline
LIFO  & Last‑in first‑out &  $\mathcal{T}(\cdot)$ & Tokenizer \\ \hline
LLM   & Large language model & $p_{\mathcal{L}}(\cdot)$ & Conditional distribution of LLM \\ \hline
LLMO  & Large language model optimizer &  $\mathbb{T}$ & Vocabulary set \\ 
\hline 
MIMO  & Multiple-input multiple-output & $\mathbb{S}$ & State space \\ \hline
RL    & Reinforcement learning & $\mathbb{S}^{\star}$/$\mathbb{S}^{\prime}$ & Optimal/non-optimal state sets \\ \hline
SE    & Spectral efficiency & $s^{(t)}$ & State at iteration $t$ \\  \hline
UE    & User equipment & $\gamma^{(t)}$ & Average convergence rate at iteration $t$ \\ \hline
\end{tabular}
\end{table*}

\section{LLM Optimizer Framework}\label{sec:sec2}

We consider an optimization problem that maximizes a reward function $r(\cdot)$. Our goal is to identify an action vector $\mathbf{x}\in\mathbb{R}^{D}$ which represents networking policies. The corresponding problem can be formulated as
\begin{align}\label{eq:P}
    &\max_{\mathbf{x}_{\sf{min}}\preceq\mathbf{x}\preceq\mathbf{x}_{\sf{max}}}\ r(\mathbf{x}),
\end{align}
where $\mathbf{x}_{\sf{min}}$ and $\mathbf{x}_{\sf{max}}$ indicate the lower and upper bounds for $\mathbf{x}$, respectively. Due to the complicated nature of wireless networks, mathematical models for $r(\cdot)$, which describes network performance indicators, are generally unavailable. Instead, reward values can be evaluated through numerical simulations or real-world measurements.

Such a black-box network management problem can be addressed by utilizing existing BBO techniques, such as the GA \cite{GA}, BO \cite{BO}, and RL methods \cite{RL}. Their effectiveness has been demonstrated in various wireless communication systems \cite{KShen:22,Yang:24,DDPG,MKim:24}. However, existing BBO algorithms require careful hyperparameter tuning, such as selection, crossover, and mutation strategies of GAs and the NN architectures of RL agents. Moreover, the resulting optimized hyperparameters are only suitable for specific reward functions, thereby limiting their generalization ability across different network scenarios.

The aforementioned challenges can be resolved via the LLMO framework \cite{OPRO,BHuang:24,LMEA,GD,LEO,ANie:23,HLee:24a,HLee:24b,MALLM,MALLM2,HLi:24,KQiu:24,JTong:24,MOEAD}. This method exploits the inherent reasoning capabilities of pretrained LLMs to find the optimal solution to \eqref{eq:P}. LLMs serve as a BBO solver which generates improved actions by observing candidate actions and their corresponding reward values without using any mathematical models of $r(\cdot)$. Unlike traditional BBO techniques, the LLMO framework does not require human intervention, such as additional fine-tuning of hyperparameters and retraining of NNs. This results in a high level of generalization, enabling an LLM to universally address various network problems. 

However, there exist no fundamental studies that establish the theoretical guarantees of the LLMO, in particular, the optimality and convergence. For this reason, existing works have been confined to developing heuristics for the LLMO that only work for specific systems. To fully understand the potential and limitations of the LLMO, we aim to investigate its fundamentals through rigorous analysis of LLM inference calculations. To this end, this section formalizes the original LLMO framework presented in \cite{OPRO}.

\subsection{LLM optimizer} 

\begin{figure}
    \centering
    \includegraphics[width=.7\linewidth]{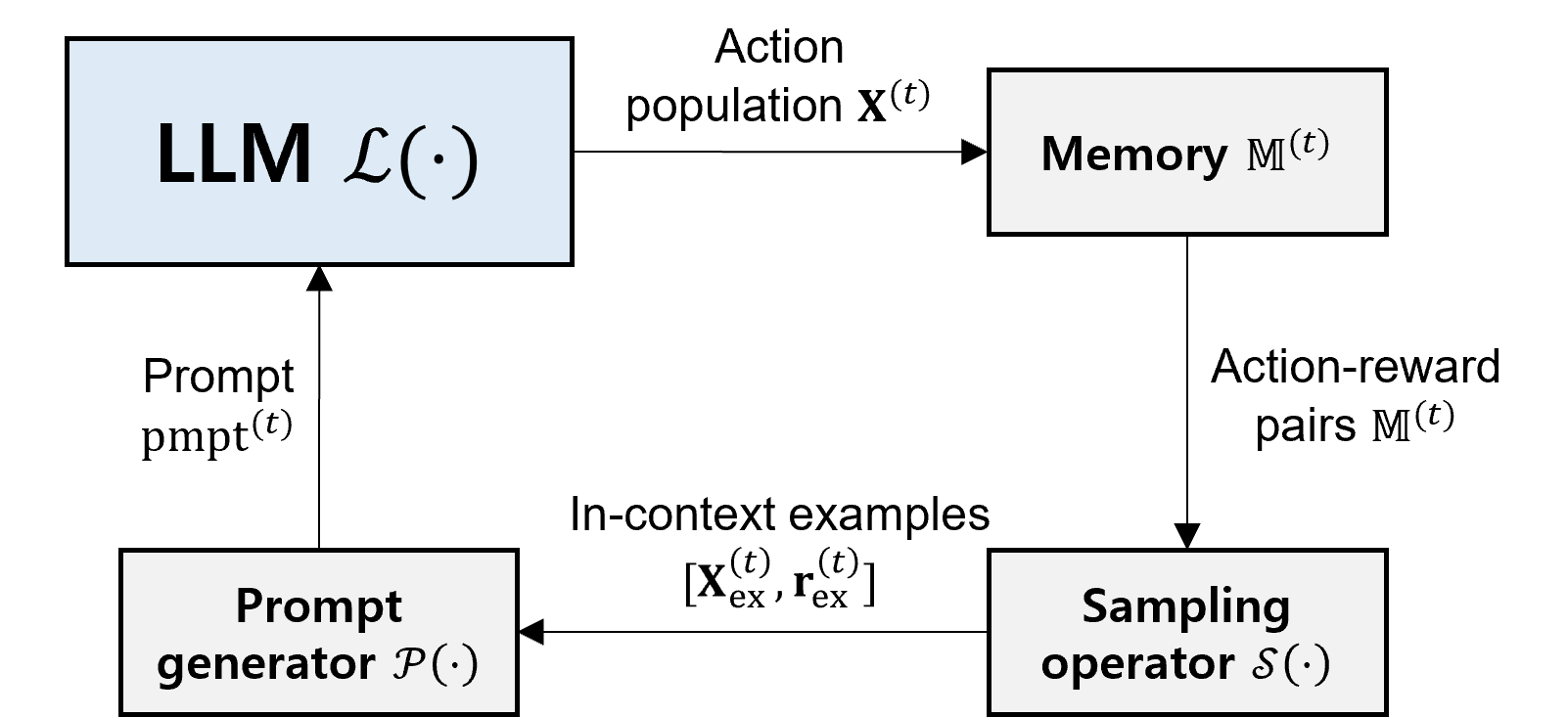}
    \caption{LLMO framework \cite{OPRO}.}
    \label{fig:LLMO}
\end{figure}

\begin{algorithm}
\caption{LLMO Framework \cite{OPRO}}
\begin{algorithmic}\label{alg:alg1}
    \STATE{Initialize $\mathbb{M}^{(0)}$, $\mathbf{x}_{\sf{best}}^{(0)}$, and $r_{\sf{best}}^{(0)}$.}
    \FOR{iteration $t=1,\cdots,T$}
        \STATE{Sample $[\mathbf{X}^{(t-1)}_{\sf{ex}},\mathbf{r}^{(t-1)}_{\sf{ex}}]$ using $\mathcal{S}(\cdot)$ in \eqref{eq:sampler}.}
        \STATE{Generate ${\sf{pmpt}}^{(t-1)}$ using $\mathcal{P}(\cdot)$ in Fig. \ref{fig:fig0}.}
        \STATE{Generate $\mathbf{X}^{(t)}$ using the LLM $\mathcal{L}(\cdot)$ in \eqref{eq:LLM}.}
        \IF{$r_{\sf{best}}^{(t-1)}<\max_{p}r(\mathbf{x}_{p}^{(t)})$}
            \STATE{Set $\mathbf{x}_{\sf{best}}^{(t)}=\max_{p}r(\mathbf{x}_{p}^{(t)})$ and $r_{\sf{best}}^{(t)}=r(\mathbf{x}^{(t)}_{\sf{best}})$.}
        \ELSE
            \STATE{Set $\mathbf{x}_{\sf{best}}^{(t)}=\mathbf{x}_{\sf{best}}^{(t-1)}$ and $r_{\sf{best}}^{(t)}=r_{\sf{best}}^{(t-1)}$.}
        \ENDIF        
        \STATE{Update the memory $\mathbb{M}^{(t)}$ using \eqref{eq:memory}.}
    \ENDFOR
    \STATE{Choose $\mathbf{x}^{(T)}_{\sf{best}}$ as the final action.}
\end{algorithmic}
\end{algorithm}

We formalize the LLMO framework originally introduced in \cite{OPRO}. As illustrated in Fig. \ref{fig:LLMO}, this method consists of pretrained LLM $\mathcal{L}(\cdot)$, memory $\mathbb{M}^{(t)}$, sampling operator $\mathcal{S}(\cdot)$, and prompt generator $\mathcal{P}(\cdot)$. Detailed procedures of the LLMO are summarized in Algorithm 1. For initialization, we randomly choose a set of $P$ actions $\mathbf{x}^{(0)}_{p} \in \mathbb{R}^{D}$ ($p = 1,\cdots,P$), collectively forming an action population matrix $\mathbf{X}^{(0)} = [\mathbf{x}^{(0)}_{1}, \cdots, \mathbf{x}^{(0)}_{P}]^{T} \in \mathbb{R}^{P \times D}$. The associated reward values are obtained as $\mathbf{r}^{(0)} = [r(\mathbf{x}^{(0)}_{1}), \cdots, r(\mathbf{x}^{(0)}_{P})]^{T} \in \mathbb{R}^{P}$. Then, the memory $\mathbb{M}^{(0)}$ is initialized as $\mathbb{M}^{(0)} = [\mathbf{X}^{(0)}, \mathbf{r}^{(0)}]$. Also, we set the best action $\mathbf{x}^{(0)}_{\sf{best}}$ and its reward $r^{(0)}_{\sf{best}}$ as $\mathbf{x}^{(0)}_{\sf{best}} = \arg\max_{p} r(\mathbf{x}_{p}^{(0)})$ and $r^{(0)}_{\sf{best}} = r(\mathbf{x}_{\sf{best}}^{(0)})$, respectively.

At the $t$-th iteration, we sample $P$ example actions $\mathbf{X}^{(t-1)}_{\sf{ex}}\in\mathbb{R}^{P\times D}$ and their rewards $\mathbf{r}^{(t-1)}_{\sf{ex}}\in\mathbb{R}^{P}$ 
from the memory $\mathbb{M}^{(t-1)}$ by using the sampling operator $\mathcal{S}(\cdot)$ as $[\mathbf{X}^{(t-1)}_{\sf{ex}},\mathbf{r}^{(t-1)}_{\sf{ex}}]=\mathcal{S}(\mathbb{M}^{(t-1)})$. 
The resulting action-reward pairs $[\mathbf{X}^{(t-1)}_{\sf{ex}},\mathbf{r}^{(t-1)}_{\sf{ex}}]$ is included in a new input prompt, which provides in-context candidate actions to the LLM. Such an ICL strategy helps align LLMs with new tasks that are not involved in the training corpus \cite{GPT3,ICL}. Thus, the LLM can extract patterns of actions with high reward values, thereby enhancing the likelihood of generating optimal actions.

In the LLMO, a selection of good sampling operators is crucial. Popular choices include an elitist sampler which extracts the top $P$ actions from the memory $\mathbb{M}^{(t-1)}$ \cite{LMEA,LEO,HLee:24a}, and a last-in-first-out (LIFO) sampler that chooses the most recent decision samples \cite{MALLM,MALLM2}. Let $[\mathbf{X}^{(t)}_{\sf{best}},\mathbf{r}_{\sf{best}}^{(t)}]$ be action-reward pairs in $\mathbb{M}^{(t)}$ corresponding to the $P$ highest reward values. Then, the output of the elitist and LIFO samplers are respectively given as
\begin{align}\label{eq:sampler}
    [\mathbf{X}^{(t-1)}_{\sf{ex}},&\mathbf{r}^{(t-1)}_{\sf{ex}}]
    \!=\!\begin{cases}
        \!\![\mathbf{X}^{(t-1)}_{\sf{best}},\mathbf{r}^{(t-1)}_{\sf{best}}], & \!\!\!\!\text{for elitist sampler},\\
        \!\![\mathbf{X}^{(t-1)},\mathbf{r}^{(t-1)}], & \!\!\!\!\text{for LIFO sampler}.
    \end{cases}
\end{align}

\begin{figure}
    \centering
    \begin{tcolorbox}[colback=blue!5!white, colframe=blue!75!black, boxrule=0.3mm]
    \textbf{Task description:} You are an agent tasked to maximize a reward function by determining D-dimensional action vector [x\_1, ..., x\_D] whose elements are between x\_min and x\_max.
    
    \textbf{Data format:} You are provided with the action-reward pairs. The first D columns stand for the action vectors, and the last column is the associated reward. \\   
    \textbf{In-context examples:}\\
    x\_1, x\_2,..., x\_D, reward\\
    0.301, 0.713,..., 0.670, 1.732\\
    ...\\
    0.415, 0.380,..., 0.827, 0.054\\    
    \textbf{Instruction:} Generate P new action vectors different from all above that can improve the reward. Actions should be presented in a CSV format of shape (P, D) where different rows indicate different action vectors. Do not generate text and codes.
    \end{tcolorbox}    
    \caption{An example prompt for LLMO.}
    \label{fig:fig0}
\end{figure}

A natural language prompt, denoted by ${\sf{pmpt}}^{(t-1)}$, is created using the in-context examples as
\begin{align}
    {\sf{pmpt}}^{(t-1)}=\mathcal{P}(\mathbf{X}^{(t-1)}_{\sf{ex}},\mathbf{r}^{(t-1)}_{\sf{ex}}), \label{eq:prompt}
\end{align}
where $\mathcal{P}(\cdot)$ represents the prompt generator shown in Fig.~\ref{fig:fig0}. The prompt consists of four different parts. The task description part outlines the optimization task details. The data format part explains the format of the in-context examples. Subsequently, the example action-reward pairs $[\mathbf{X}^{(t-1)}_{\sf{ex}}, \mathbf{r}^{(t-1)}_{\sf{ex}}]$ are provided in a comma-separated value (CSV) format. Finally, the instruction part guides the LLM to produce $P$ new actions.

The output of the LLM $\mathcal{L}(\cdot)$ can be expressed as
\begin{align}\label{eq:LLM}
    \mathbf{X}^{(t)}=[\mathbf{x}^{(t)}_{1},\cdots,\mathbf{x}^{(t)}_{P}]^{T}=\mathcal{L}({\sf{pmpt}}^{(t-1)}),
\end{align}
where the output action population $\mathbf{X}^{(t)}$ consists of $P$ new actions $\mathbf{x}^{(t)}_{p}$ ($p=1,\cdots,P$) with corresponding reward values $\mathbf{r}^{(t)}=[r(\mathbf{x}^{(t)}_{1}),\cdots,r(\mathbf{x}^{(t)}_{P})]^{T}$. If the new actions achieve a better reward than $r_{\sf{best}}^{(t-1)}$, we update $\mathbf{x}_{\sf{best}}^{(t)}$ and $r_{\sf{best}}^{(t)}$ accordingly. Also, the new decisions are added to the memory $\mathbb{M}^{(t-1)}$. According to the sampling operators in \eqref{eq:sampler}, instead of accumulating all decision history, 
it suffices to retain the new samples $[\mathbf{X}^{(t)},\mathbf{r}^{(t)}]$ and current in-context examples $[\mathbf{X}^{(t-1)}_{\sf{ex}},\mathbf{r}^{(t-1)}_{\sf{ex}}]$ in the memory $\mathbb{M}^{(t)}$ as
\begin{align}\label{eq:memory}
    \mathbb{M}^{(t)}=\begin{bmatrix}
        \mathbf{X}^{(t)} & \mathbf{r}^{(t)}\\
        \mathbf{X}^{(t-1)}_{\sf{ex}} & \mathbf{r}^{(t-1)}_{\sf{ex}}
    \end{bmatrix}.
\end{align}
These procedures are repeated until the predefined maximum iteration number $T$. Finally, $\mathbf{x}^{(T)}_{\sf{best}}$ is selected as a solution.

\subsection{Implementation in wireless networks}

\begin{figure}
    \centering
    \includegraphics[width=.7\linewidth]{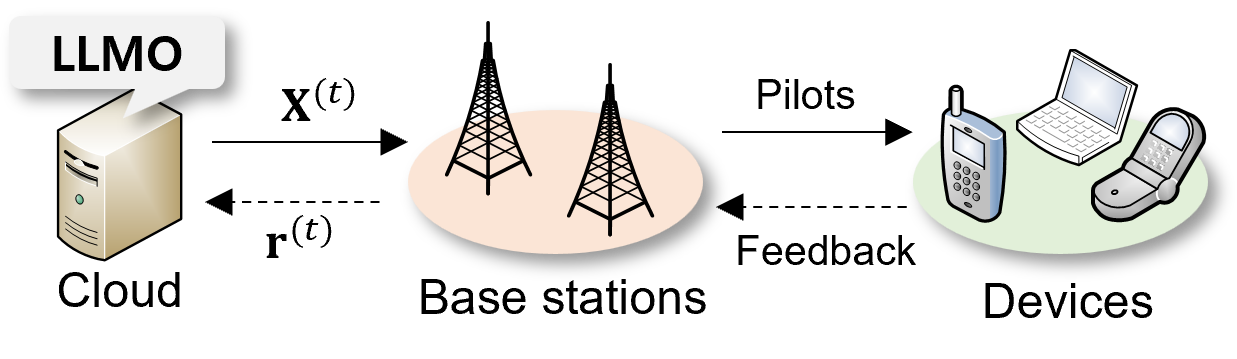}
    \caption{Implementation of LLMO in wireless networks.}
    \label{fig:imp}
\end{figure}

Fig. \ref{fig:imp} illustrates the practical implementation of the LLMO in wireless networks with black-box reward functions. A GPU‑enabled cloud server handles all computations of the LLMO including LLM inference, memory management, sampling, and prompt generation. To this end, it is essential to evaluate rewards associated with the new action population. However, due to the nature of the BBO, the cloud cannot quantify reward values by itself.

Similar to traditional RL methods, this can be achieved through interaction with network entities such as base stations (BSs) and wireless devices. The cloud first determines the new action population $\mathbf{X}^{(t)}$, which describes transmission policies of BSs, e.g., transmit power levels and resource allocation strategies. It is then conveyed to the BSs through backhaul links. To obtain the associated reward values $\mathbf{r}^{(t)}$, the BSs transmit pilot signals based on $\mathbf{X}^{(t)}$ to devices through wireless access links. Then, the devices measure key metrics such as receive signal strength, interference power, and achievable data rate. Through reliable feedback channels, such information is sent back to the BSs, which is consolidated into the reward values $\mathbf{r}^{(t)}$. Finally, the cloud collects them into the memory along with the action population $\mathbf{X}^{(t)}$. Consequently, the LLMO can be deployed in practical wireless networks.

Implementing the LLMO incurs coordination overheads among network entities, in particular, pilot transmission and feedback through wireless access links. This is closely related to the number of iterations $T$ of Algorithm \ref{alg:alg1}. As $T$ grows, the LLMO can enhance the network performance at the expense of the increased overhead. Therefore, to choose a proper $T$ that balances performance and overhead, we need to investigate the convergence behavior of the LLMO framework.

Several prior works evaluated the LLMO in various wireless communication networks with black-box reward functions \cite{HLee:24a,HLee:24b,MALLM,MALLM2,HLi:24,JTong:24,KQiu:24}. The effectiveness of the LLMO has been demonstrated through numerical simulations, but these are limited to specific scenarios. Thus, the universal applicability of the LLMO for handling generic black-box network management tasks has not yet been verified. To this end, it is necessary to establish the theoretical foundation of the LLMO, in particular, its convergence and optimality. 

\subsection{Relationships with prompt engineering techniques}


The LLMO can be viewed as a prompt engineering approach that efficiently provides instructions for desired tasks, enabling LLMs to accurately address various reasoning problems \cite{GPT3,ICL,CoT,CoT-SC}. The LLMO adopts the ICL prompting strategy that leverages the action-reward pairs as in-context examples. Thanks to their remarkable reasoning ability, LLMs can discern patterns and information from prompts and generate appropriate responses without fine-tuning \cite{GPT3,ICL}.

The ICL strategy, however, might not be suitable for tackling complicated tasks that require multiple reasoning steps. This can be addressed by the chain-of-thought (CoT) technique, which breaks down complex problems into several decision steps called thoughts \cite{CoT,CoT-SC}. Prompting a sequence of thoughts guides LLMs to generate intermediate reasoning steps toward the final answer. In the LLMO, the in-context examples $\mathbf{X}_{\sf{ex}}^{(t)}$, which contain the decision history of the LLM, serve as thoughts to reach the globally optimal action. 

Unlike traditional CoT approaches that rely on man-made reasoning steps, the LLMO autonomously yields thoughts starting from the initial action population $\mathbf{X}^{(0)}$. This can be viewed as a variation of the zero-shot CoT technique \cite{ZCoT} which samples initial thoughts using LLMs and feeds the resulting thoughts back. Such a self-feedback mechanism has been proven to be efficient to improve the reasoning capability of LLMs without any human intervention \cite{Reflextion}. 

These prompt engineering methods suffer from the uncertainty in LLM responses, exemplified by the hallucination issue \cite{Hallucination,Hallucination2}, where LLMs generate unfaithful or nonsensical content. In the LLMO framework, the LLM hallucination produces poor action populations that incur degraded reward values \cite{OPRO}. For this reason, mitigating the hallucination has been a crucial feature of the LLMO. Nevertheless, the LLMO has been numerically shown to reach the global optimum in various BBO tasks \cite{BHuang:24,LMEA,GD,ANie:23,LEO,MOEAD,HLee:24a,HLee:24b,MALLM,MALLM2,HLi:24,KQiu:24,JTong:24}. This motivates our theoretical study of the LLMO framework in the following sections.

\section{Theoretical Analysis of LLM Optimizer}\label{sec:sec3}

Despite various interpretations and extensive numerical demonstrations, the true capability of the LLMO has not been adequately explored yet, especially regarding its optimality for the BBO problems. This section presents a theoretical analysis and reveals the core features of the LLMO. We begin by examining the inference computations of the LLM, which is followed by a detailed theoretical analysis.

\subsection{LLM inference}

The LLM inference consists of encoding, embedding, transformer layers, and decoding steps \cite{GPT2}. In the encoding step, natural language inputs are pre-processed by using a tokenizer $\mathcal{T}(\cdot)$, which divides an input prompt ${\sf{pmpt}}$ into smaller units called tokens, each containing several characters or subwords. Each token is then converted into its corresponding integer token identification (ID). Thus, the tokenizer is viewed as a lookup table that employs a pretrained vocabulary $\mathbb{Z}$ collecting all possible tokens. For the remainder of this paper, we adopt the terms tokens and token IDs interchangeably. The token vector for the input prompt $\sf{pmpt}$ is expressed as
\begin{align}
    \mathbf{z}^{\sf{in}}=[z_{1}^{\sf{in}},\cdots,z_{N_{\sf{in}}}^{\sf{in}}]=\mathcal{T}({\sf{pmpt}}),
\end{align}
where $z_{i}^{\sf{in}}\in\mathbb{Z}$ is the $i$-th token of the input prompt and $N_{\sf{in}}$ stands for the number of tokens. 
Next, the embedding layer $\mathcal{E}(\cdot)$ is applied to transform each token $z_{i}^{\sf{in}}$ into an embedding vector $\mathbf{e}_{i}^{\sf{in}}$ as $\mathbf{e}_{i}^{\sf{in}} = \mathcal{E}(z_{i}^{\sf{in}})$. The collection of embeddings $\mathbf{e}^{\sf{in}} \triangleq [\mathbf{e}_{1}^{\sf{in}}, \cdots, \mathbf{e}_{N_{\sf{in}}}^{\sf{in}}]$ is then processed by transformer layers $\mathcal{A}(\cdot)$, which consist of multi-head masked self-attention, feed-forward networks, and layer normalizations \cite{GPT3,LLaMA}.

Current LLMs rely on the autoregressive architecture that determines output tokens sequentially. Let $z^{\sf{out}}_{k}$ be the $k$-th output token. Given previous output tokens $\mathbf{z}_{[1:k-1]}^{\sf{out}}\triangleq[z_{1}^{\sf{out}},\cdots,z_{k-1}^{\sf{out}}]$, we obtain the logit vector $\mathbf{b}_{k}$ for the $k$-th output token $z^{\sf{out}}_{k}$ using the transformer layers $\mathcal{A}(\cdot)$ as
\begin{align}
    \mathbf{b}_{k}=\mathcal{A}([\mathbf{e}^{\sf{in}},\mathbf{e}_{1}^{\sf{out}},\cdots,\mathbf{e}_{k-1}^{\sf{out}}]),
\end{align}
where $\mathbf{e}_{k}^{\sf{out}}\triangleq\mathcal{E}(z^{\sf{out}}_{k})$ accounts for the embedding vector of the $k$-th output token $z^{\sf{out}}_{k}$. 
Let $p_{\mathcal{L}}(\cdot|\cdot)$ be a learned conditional distribution of the LLM $\mathcal{L}(\cdot)$. Then, the probability of generating $z_{k}^{\sf{out}}\in\mathbb{Z}$ is obtained using the softmax function $\sigma(\cdot)$~as
\begin{align}\label{eq:softmax}
    p_{\mathcal{L}}(z_{k}^{\sf{out}}|\mathbf{z}_{[1:k-1]}^{\sf{out}},\mathbf{z}^{\sf{in}})\!=\![\sigma(\mathbf{b}_{k})]_{z_{k}^{\sf{out}}}\!=\!\frac{\exp([\mathbf{b}_{k}]_{z_{k}^{\sf{out}}}/\alpha) }{\sum_{z\in\mathbb{Z}}\exp([\mathbf{b}_{k}]_{z}/\alpha)},
\end{align}
where $[\mathbf{u}]_{v}$ represents the $v$-th element of a vector $\mathbf{u}$, and $\alpha > 0$ is the temperature parameter of the LLM controlling the creativity of output responses \cite{CoT-SC}.

To further enhance the creativity, LLMs employ random sampling strategies, such as nucleus, top-K, and temperature sampling \cite{GPT2,CoT-SC}. For instance, in the nucleus and top-K sampling methods, the LLM first calculates \eqref{eq:softmax} and constructs a restricted vocabulary set $\mathbb{T} \subset \mathbb{Z}$ by selecting tokens with the top probabilities. Thus,  $p_{\mathcal{L}}(z_{k}^{\sf{out}}|\mathbf{z}_{[1:k-1]}^{\sf{out}},\mathbf{z}^{\sf{in}})$ is refined as
\begin{align}\label{eq:topk}
    p_{\mathcal{L}}(z_{k}^{\sf{out}}|\mathbf{z}_{[1:k-1]}^{\sf{out}},\mathbf{z}^{\sf{in}})=\!\!
    \begin{cases}
    \frac{\exp([\mathbf{b}_{k}]_{z_{k}^{\sf{out}}}/\alpha) }{\sum_{z\in\mathbb{T}}\exp([\mathbf{b}_{k}]_{z}/\alpha)}, &  \text{for}\ z_{k}^{\sf{out}}\!\in\!\mathbb{T},\\
    0, & \text{for}\ z_{k}^{\sf{out}}\!\notin\!\mathbb{T}.
    \end{cases}
\end{align}
The output tokens are then randomly sampled from $\mathbb{T}$ according to the distribution in \eqref{eq:topk}.

Let $N_{\sf{out}}$ be the number of output tokens. Then, the probability of the output token vector $\mathbf{z}^{\sf{out}}\triangleq[z_{1}^{\sf{out}},\cdots,z_{N_{\sf{out}}}^{\sf{out}}]$ for a given $\mathbf{z}^{\sf{in}}$ is obtained~as
\begin{align}
    p_{\mathcal{L}}(\mathbf{z}^{\sf{out}}|\mathbf{z}^{\sf{in}})=\prod_{k=1}^{N_{\sf{out}}}p_{\mathcal{L}}(z_{k}^{\sf{out}}|\mathbf{z}_{[1:k-1]}^{\sf{out}},\mathbf{z}^{\sf{in}}). \label{eq:zout}
\end{align}
Finally, we can identify natural language outputs using the decoding process $\mathcal{T}^{-1}(\cdot)$.
Consequently, the LLM inference \eqref{eq:LLM} of the LLMO can be rewritten by
\begin{align}\label{eq:LLM2}
    \mathbf{X}^{(t)}=\mathcal{T}^{-1}\Big(\sigma\big(\mathcal{A}\big(\mathcal{E}(\mathcal{T}({\sf{pmpt}}^{(t-1)}))\big)\big)\Big).
\end{align}

\subsection{LLMO as finite-state Markov chain}\label{sec:sec3b}

\begin{figure*}
    \centering
    \includegraphics[width=\linewidth]{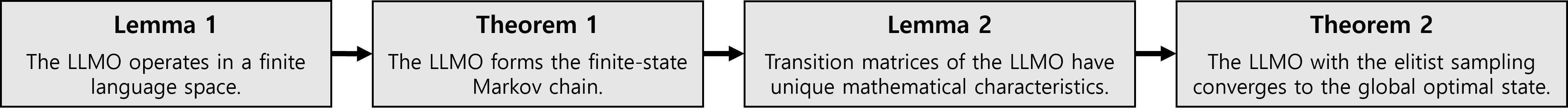}
    \caption{Outline of convergence analysis.}
    \label{fig:fig3m}
\end{figure*}

Based on the analysis of the LLM inference, we will prove the convergence of the LLMO framework. Fig. \ref{fig:fig3m} illustrates our proof strategies. This subsection interprets the LLMO framework as a finite-space Markov chain. To this end, Lemma \ref{prop:prop1} first investigates a state space of the LLMO. It is followed by Theorem \ref{prop:prop3} which shows the Markovian property of the LLMO. Next, Lemma \ref{prop:prop4} reveals unique features of transition matrices of the LLMO, which lead to the rigorous convergence proof in Theorem \ref{prop:prop5}.

To analyze the LLMO, we establish the following two assumptions.
\begin{assumption}\label{amp:amp1}
The vocabulary set $\mathbb{T}$ of the LLMO contains tokens relevant for generating the action population matrix $\mathbf{X}^{(t)}$ in a CSV format. 
\end{assumption}
\begin{assumption}\label{amp:amp2}
The LLM generates elements of $\mathbf{X}^{(t)}$ as finite-precision floating-point numbers with $N_{\sf{digit}}$ digits. 
\end{assumption}

Provided that the LLM grasps the instruction part of the input prompt in Fig. \ref{fig:fig0}, tokens which represent numbers in $\mathbf{X}^{(t)}$ will have high probabilities and be included in $\mathbb{T}$ according to \eqref{eq:topk}. Therefore, Assumption \ref{amp:amp1} holds in general. Also, current LLMs are implemented with weight matrices and bias vectors consisting of finite-precision floating-point numbers. Network management tasks typically impose a feasible set bounded by $\mathbf{x}_{\sf{min}}$ and $\mathbf{x}_{\sf{max}}$. Hence, actions obtained by the LLM can be interpreted with finite digits.



The following lemma discusses the tokenizer in the LLMO, particularly focusing on the byte-pair encoding (BPE) tokenizer $\mathcal{T}_{\sf{BPE}}(\cdot)$ used in GPT-3.5-Turbo. Notice that the BPE tokenizer and its variants have been widely adopted in various LLMs including LLaMA3, Mistral, and DeepSeek.

\begin{lemma}\label{prop:prop1}
A set of all token vectors $\mathbb{S}\triangleq\{\mathcal{T}_{\sf{BPE}}(\mathbf{X}):\forall \mathbf{X}\in\mathbb{R}^{P\times D}\}$ associated with any matrix $\mathbf{X}=[\mathbf{x}_{1},\cdots,\mathbf{x}_{P}]^{T}$ containing $P$ action vectors $\mathbf{x}_{p}$ for $p=1,\cdots,P$ is given as
\begin{align}
    \mathbb{S}=\mathbb{T}^{PD N_{\sf{token}}}, \label{eq:S}
\end{align}
where $\mathbb{U}^{V}$ stands for a $V$-ary Cartesian product of a set $\mathbb{U}$, $N_{\sf{token}}\triangleq\lceil N_{\sf{digit}}/3\rceil + 3$ stands for the number of tokens per each floating-point number, and the vocabulary set $\mathbb{T}$ of the LLMO is defined as
\begin{align}\label{eq:Z}
        \mathbb{T}\!=\!\{\!z_{\sf{0}},\!\cdots\!,z_{\sf{9}},z_{\sf{00}},\!\cdots\!,z_{\sf{99}},z_{\sf{000}},\!\cdots\!,z_{\sf{999}},z_{\sf{.}},z_{\sf{-}},z_{\sf{,}},z_{\sf{\backslash n}}\!\}
\end{align}
with $z_{\sf{str}}\triangleq\mathcal{T}_{\sf{BPE}}({\sf{str}})$ being a token of a string ${\sf{str}}$.
\end{lemma}

\begin{IEEEproof}
See Appendix A. 
\end{IEEEproof}

Lemma \ref{prop:prop1} reveals that the LLMO with the BPE tokenizer operates in the finite language space $\mathbb{S}$ in \eqref{eq:S}. This means that any action population $\mathbf{X}^{(t)}$ as well as in-context examples $\mathbf{X}^{(t)}_{\sf{ex}}$ can be interpreted as a discrete representation within $\mathbb{S}$. Deriving $\mathbb{S}$ for other tokenizers is not straightforward. In this case, the vocabulary set $\mathbb{T}$ might contain additional tokens, which potentially increase the number of tokens $N_{\sf{token}}$. Nevertheless, the cardinality $|\mathbb{T}|$ and $N_{\sf{token}}$ are finite numbers, indicating that the LLMO realized with arbitrary tokenizers still relies on a finite language space $\mathbb{S}$. Therefore, our subsequent analysis holds for diverse types of LLMs and tokenizers.

Based on these results, we can model the LLMO, including the LLM inference $\mathcal{L}(\cdot)$ in \eqref{eq:LLM2} and the sampling operator $\mathcal{S}(\cdot)$ in \eqref{eq:sampler}, as a finite-state Markov chain as follows.

\begin{theorem}\label{prop:prop3}
    The LLMO forms the Markov chain as
    \begin{align}\label{eq:markov}
        \mathbf{X}^{(0)}_{\sf{ex}}\rightarrow\mathbf{X}^{(1)}_{\sf{ex}}\rightarrow\cdots\rightarrow\mathbf{X}^{(T)}_{\sf{ex}},
    \end{align}
    where $T$ stands for the maximum iteration number of Algorithm \ref{alg:alg1} and $\mathcal{T}_{\sf{BPE}}(\mathbf{X}^{(t)}_{\sf{ex}})\in\mathbb{S}$. The transition probability $\Pr\{\mathbf{X}^{(t)}_{\sf{ex}}|\mathbf{X}^{(t-1)}_{\sf{ex}}\}$ is computed by
        \begin{align}\label{eq:trans_prob}
        \Pr\{\mathbf{X}^{(t)}_{\sf{ex}}&|\mathbf{X}^{(t-1)}_{\sf{ex}}\}=\!\!\!\!\!\!\!\!\!\sum_{\mathcal{T}_{\sf{BPE}}(\mathbf{X}^{(t)})\in\mathbb{S}}\!\!\!\!\Big(p_{\mathcal{S}}(\mathbf{X}^{(t)}_{\sf{ex}}|\mathbf{X}^{(t)},\mathbf{X}^{(t-1)}_{\sf{ex}})\\
    &\times p_{\mathcal{L}}(\mathcal{T}_{\sf{BPE}}(\mathbf{X}^{(t)})|\mathcal{T}_{\sf{BPE}}(\mathbf{X}^{(t-1)}_{\sf{ex}}),\mathcal{T}_{\sf{BPE}}(\mathbf{r}^{(t-1)}_{\sf{ex}}),\mathbf{z}_{\sf{fix}})\Big),\nonumber
    \end{align}
    where $p_{\mathcal{S}}(\cdot|\cdot)$ accounts for the transition probability of the sampling operator $\mathcal{S}(\cdot)$ in \eqref{eq:sampler} and $\mathbf{z}_{\sf{fix}}$ is the fixed token vector of $\sf{pmpt}^{(t-1)}$ excluding the in-context example part.
\end{theorem}
\begin{IEEEproof}
See Appendix B. 
\end{IEEEproof}

Theorem \ref{prop:prop3} states that the action population matrices of the in-context examples $\mathbf{X}_{\sf{ex}}^{(t)}$ form a Markov chain within the finite language space $\mathcal{T}_{\sf{BPE}}(\mathbf{X}_{\sf{ex}}^{(t)})\in\mathbb{S}$. 
The transition probability in \eqref{eq:trans_prob} captures the effects of both the sampling operator $\mathcal{S}(\cdot)$ and the random token sampling in \eqref{eq:topk}. Thus, our analysis explicitly involves unpredictable LLM responses including the hallucination issue. Notice that the result in \eqref{eq:trans_prob} is valid for any stochastic sampling operators $p_{\mathcal{S}}(\mathbf{X}^{(t)}_{\sf{ex}}|\mathbf{X}^{(t)},\mathbf{X}^{(t-1)}_{\sf{ex}})$.

Since the LLMO continually updates the best action $\mathbf{x}_{\sf{best}}^{(t)}$ based on $\mathbf{X}^{(t)}_{\sf{ex}}$, it suffices to prove the optimality of the LLMO through the Markov chain in \eqref{eq:markov}. In the following, we will show that the LLMO with the elitist sampler converges to the optimal action. This provides the theoretical foundations for existing works \cite{OPRO,BHuang:24,LMEA,GD,LEO,ANie:23,HLee:24a,HLee:24b,MALLM,MALLM2,HLi:24,KQiu:24,JTong:24,MOEAD}.

\subsection{Convergence analysis} \label{sec:sec3c}

To proceed with the convergence analysis, we introduce essential definitions in the following.
\begin{definition}\label{def:def1}
A token vector $\mathcal{T}_{\sf{BPE}}(\mathbf{X}_{\sf{ex}}^{(t)})$ of the in-context examples $\mathbf{X}_{\sf{ex}}^{(t)}$ is represented as a state $s^{(t)}\triangleq\mathcal{T}_{\sf{BPE}}(\mathbf{X}_{\sf{ex}}^{(t)})$ in the finite state space $s^{(t)}\in\mathbb{S}$.
Conversely, the in-context examples are retrieved through the decoding process as $\mathbf{X}_{\sf{ex}}^{(t)}=\mathcal{T}_{\sf{BPE}}^{-1}(s^{(t)})$.
\end{definition}
\begin{definition}\label{def:def3}
A state $s^{(t)}\in\mathbb{S}$ is said to be an optimal state if the associated action population $\mathbf{X}_{\sf{ex}}^{(t)}$ contains at least one globally optimal action to the problem in \eqref{eq:P}.
\end{definition}
\begin{definition}\label{def:def4}
The optimal state set $\mathbb{S}^{\star}\subset\mathbb{S}$ is defined as a set of all optimal states. Also, $\mathbb{S}^{\prime}=\mathbb{S}\setminus\mathbb{S}^{\star}$ denotes the set of non-optimal states.
\end{definition}
\begin{definition}\label{def:def5}
An inequality $s\succ\tilde{s}$ defines the ordering between two states $s,\tilde{s}\in\mathbb{S}$ if the maximum reward among $P$ actions in $\mathcal{T}_{\sf{BPE}}^{-1}(s)$ is greater than that of $\mathcal{T}_{\sf{BPE}}^{-1}(\tilde{s})$. If they have the same maximum reward, we proceed by comparing the next largest rewards. Two states are said to be equivalent, i.e., $s= \tilde{s}$, if all rewards are identical.
\end{definition}
\begin{definition}\label{def:def7}
Without loss of generality, states are assumed to be sorted in descending order in $\mathbb{S}=\{s_{1},s_{2},\cdots,s_{|\mathbb{S}|}\}$ such that $s_{1}\succeq s_{2}\cdots\succeq s_{|\mathbb{S}|}$. Thus, the first $|\mathbb{S}^{\star}|$ states $s_{1},\cdots,s_{|\mathbb{S}^{\star}|}$ correspond to the optimal states, while the remaining $|\mathbb{S}^{\prime}|=|\mathbb{S}|-|\mathbb{S}^{\star}|$ states belong to the non-optimal states.
\end{definition}

According to the above definitions, we can define the transition probability matrix of the LLMO $\mathbf{P}_{\sf{LLM}}\in\mathbb{R}^{|\mathbb{S}|\times|\mathbb{S}|}$ which collects the transition probabilities $p_{s\tilde{s}}\triangleq\Pr\{s^{(t)}=s|s^{(t-1)}=\tilde{s}\}$ as its $(s,\tilde{s})$-th element. It is given by
\begin{align}\label{eq:pij}
    p_{s\tilde{s}}=\Pr\{\mathbf{X}^{(t)}_{\sf{ex}}=\mathcal{T}^{-1}_{\sf{BPE}}(s)|\mathbf{X}^{(t-1)}_{\sf{ex}}=\mathcal{T}^{-1}_{\sf{BPE}}(\tilde{s})\}.
\end{align}
Since the states are sorted in the descending order, $\mathbf{P}_{\sf{LLM}}$ can be expressed~as
\begin{align}\label{eq:P_LLM}
    \mathbf{P}_{\sf{LLM}}=\begin{bmatrix}
        \mathbf{P}_{1} & \mathbf{P}_{2} \\
        \mathbf{P}_{3} & \mathbf{P}_{4}
    \end{bmatrix},
\end{align}
where $\mathbf{P}_{1}\triangleq\{p_{s\tilde{s}}$:$\forall s,\tilde{s}\in\mathbb{S}^{\star}\}\in\mathbb{R}^{|\mathbb{S}^{\star}|\times|\mathbb{S}^{\star}|}$ and $\mathbf{P}_{4}\triangleq\{p_{s\tilde{s}}$:$\forall s,\tilde{s}\in\mathbb{S}^{\prime}\}\in\mathbb{R}^{|\mathbb{S}^{\prime}|\times|\mathbb{S}^{\prime}|}$ indicate the transition matrices within the optimal states $\mathbb{S}^{\star}$ and non-optimal states $\mathbb{S}^{\prime}=\mathbb{S}\setminus\mathbb{S}^{\star}$, respectively, and the matrices $\mathbf{P}_{2}\triangleq\{p_{s\tilde{s}}$:$\forall s\in\mathbb{S}^{\star},\forall \tilde{s}\in\mathbb{S}^{\prime}\}\in\mathbb{R}^{|\mathbb{S}^{\star}|\times|\mathbb{S}^{\prime}|}$ and $\mathbf{P}_{3}\triangleq\{p_{s\tilde{s}}$:$\forall s\in\mathbb{S}^{\prime},\forall \tilde{s}\in\mathbb{S}^{\star}\}\in\mathbb{R}^{|\mathbb{S}^{\prime}|\times|\mathbb{S}^{\star}|}$ are obtained as the transition probabilities from non-optimal to optimal states and optimal to non-optimal states, respectively. 

\begin{figure}
    \centering
    \includegraphics[width=.5\linewidth]{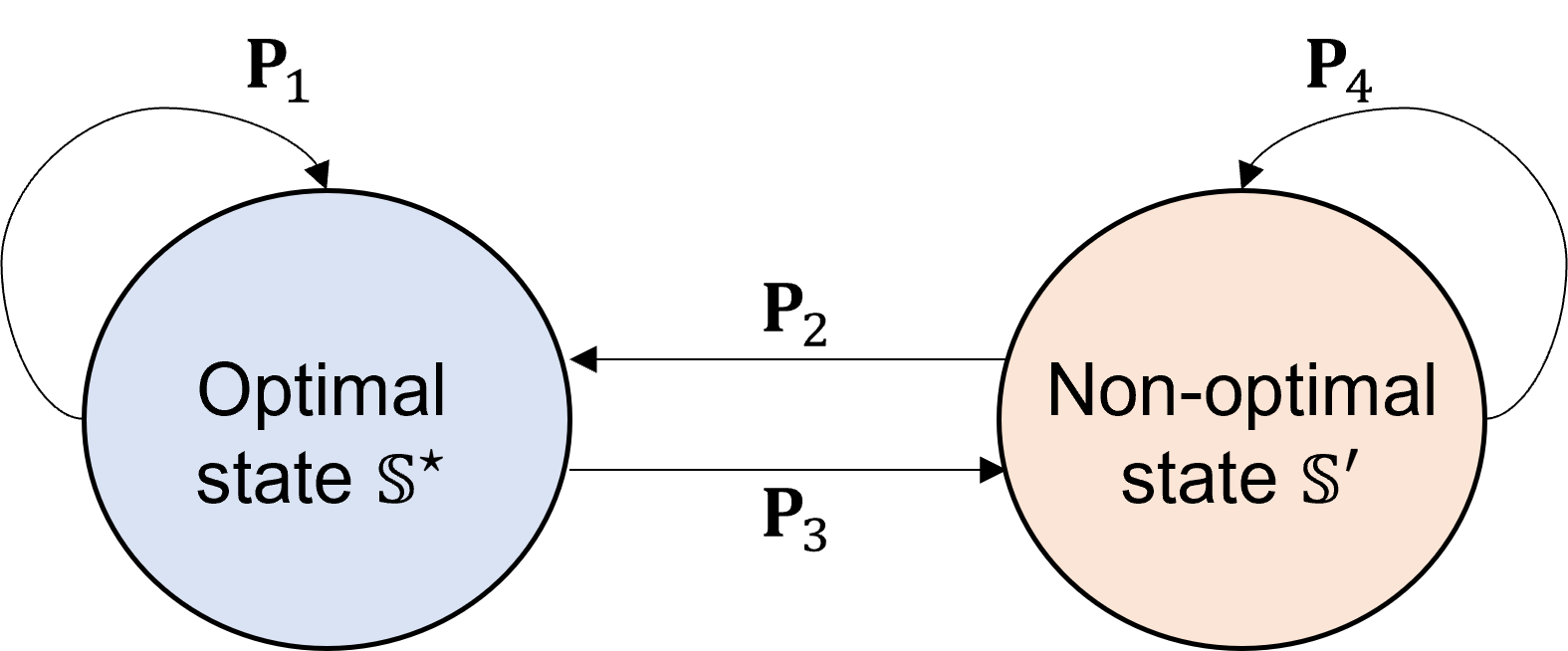}
    \caption{Interpretation of transition matrices.}
    \label{fig:fig4m}
\end{figure}

Fig. \ref{fig:fig4m} illustrates relationships among four transition matrices $\mathbf{P}_{1},\cdots,\mathbf{P}_{4}$. The LLMO can be interpreted as a two-state Markov chain consisting of the optimal state space $\mathbb{S}^{\star}$ and the non-optimal space $\mathbb{S}^{\prime}$. The LLMO changes from the non-optimal state to the optimal one with probability $\mathbf{P}_{2}$, whereas it stays in the non-optimal state with probability $\mathbf{P}_{4}$. It can also switch back from the optimal to non-optimal states with probability $\mathbf{P}_{3}$. Consequently, the convergence behavior of the LLMO is governed by two transition matrices $\mathbf{P}_{3}$ and $\mathbf{P}_{4}$. These probabilities become dominant when the LLM produces uncertain responses, i.e., poor action populations, due to the hallucination. Mitigating the LLM hallucination is challenging, and thus reducing $\mathbf{P}_{3}$ and $\mathbf{P}_{4}$ is not straightforward.

We can handle this issue via carefully designed sampling distribution $p_{\mathcal{S}}(\cdot|\cdot)$. When $\mathbf{P}_{2}$ has at least one positive entry in each column, the eigenvalue of $\mathbf{P}_{4}$ should be smaller than unity. As a result, the impact of the transition $\mathbf{P}_{4}$ becomes negligible as the number of iterations grows, and the LLMO initialized with any non-optimal actions can generate the optimal solution. Also, if $\mathbf{P}_{3}=\mathbf{0}_{|\mathbb{S}^{\prime}|\times|\mathbb{S}^{\star}|}$, then there exists no transition from the optimal to non-optimal states, thereby ensuring the optimality. If the LLMO satisfies these properties, we can prove that the LLMO always converges to the optimal action for any given initial action population $\mathbf{X}^{(0)}$. In the following lemma, we show that the LLMO with the elitist sampling indeed secures such mathematical properties.

\begin{lemma}\label{prop:prop4}
With the elitist sampling, $\mathbf{P}_{1}$ and $\mathbf{P}_{4}$ become upper triangular, $\mathbf{P}_{2}$ has at least one positive element at each column, and $\mathbf{P}_{3}=\mathbf{0}_{|\mathbb{S}^{\prime}|\times|\mathbb{S}^{\star}|}$. In contrast, for the LIFO sampling, all matrices $\mathbf{P}_{1},\cdots,\mathbf{P}_{4}$ have positive elements.
\end{lemma}
\begin{IEEEproof}
See Appendix C. 
\end{IEEEproof}

Lemma \ref{prop:prop4} reveals crucial features of the elitist sampling. The upper triangular $\mathbf{P}_{4}$ implies that the reward performance of the LLMO is non-decreasing over iterations. Also, due to the property of $\mathbf{P}_{2}$, the LLMO can move from any non-optimal state to the optimal state after several iterations. Also, the elitist operator forces $\mathbf{P}_{3}$ to a zero matrix. This guarantees that the LLMO converges to the optimum since it retains the optimal states and prevents transitions to non-optimal ones. However, such a characteristic cannot be met for the LIFO sampler, as it potentially has a nonzero $\mathbf{P}_{3}$. These intuitions can be rigorously verified in the following theorem.

\begin{theorem}\label{prop:prop5}
Suppose an arbitrary initial distribution $\Pr\{s^{(0)}=s\}>0$, $\forall s\in\mathbb{S}$. With the elitist sampler, the LLMO converges to the optimal state almost surely, i.e., $\lim_{t\rightarrow\infty}\Pr\{s^{(t)}\in\mathbb{S}^{\star}\}=1$. In contrast, the LIFO sampler is not guaranteed to identify the optimal action as $\lim_{t\rightarrow\infty}\Pr\{s^{(t)}\in\mathbb{S}^{\star}\}<1$.
\end{theorem}
\begin{IEEEproof}
See Appendix D. 
\end{IEEEproof}

From Theorem \ref{prop:prop5}, we can conclude that the elitist sampling, which chooses the $P$ best actions in the memory and prompts them to the LLM as the in-context examples, is essential for ensuring convergence. Notably, the optimality holds regardless of the convexity of the reward function or the initial distribution, demonstrating the universality of the LLMO to any given BBO task. Notice that the elitist sampling has been widely utilized in existing LLMO methods \cite{OPRO,LEO,LMEA,HLee:24a}, but its superiority over other sampling strategies has been demonstrated only through simulations. 
Theorem~\ref{prop:prop5} is the first attempt to analyze the optimality of these approaches.

The elitist sampling and its variants have been extensively adopted in the GA \cite{GRudolph:94,JSuzuki:95,HJun:16}. 
Unlike the GA which involves simple stochastic operations, the analysis of the LLMO requires a careful integration of the fundamental features of LLMs. It should be noted that our analysis includes the built-in functionalities of state-of-the-art LLMs, such as tokenizer, random token sampling, and in-context example generation, which is not straightforward compared to \cite{GRudolph:94,JSuzuki:95,HJun:16}.

\section{Multi-LLM Optimizer Framework}\label{sec:sec4}
Section \ref{sec:sec3} has proved that the LLMO attains the optimum as $t\rightarrow\infty$. Thus, even with elitist sampling, the LLMO might suffer from slow convergence, as reported in various application scenarios \cite{OPRO,LEO,LMEA,BHuang:24,HLee:24a}. This drawback can be explained via the hallucination \cite{OPRO,LEO,HLee:24a}. Since the LLMO maintains the memory that stores past decisions, hallucinated outputs with low reward values populate the memory with non-optimal in-context examples. Consequently, the LLMO gets stuck to poor actions and converges slowly.


\begin{figure}
    \centering
    \includegraphics[width=.7\linewidth]{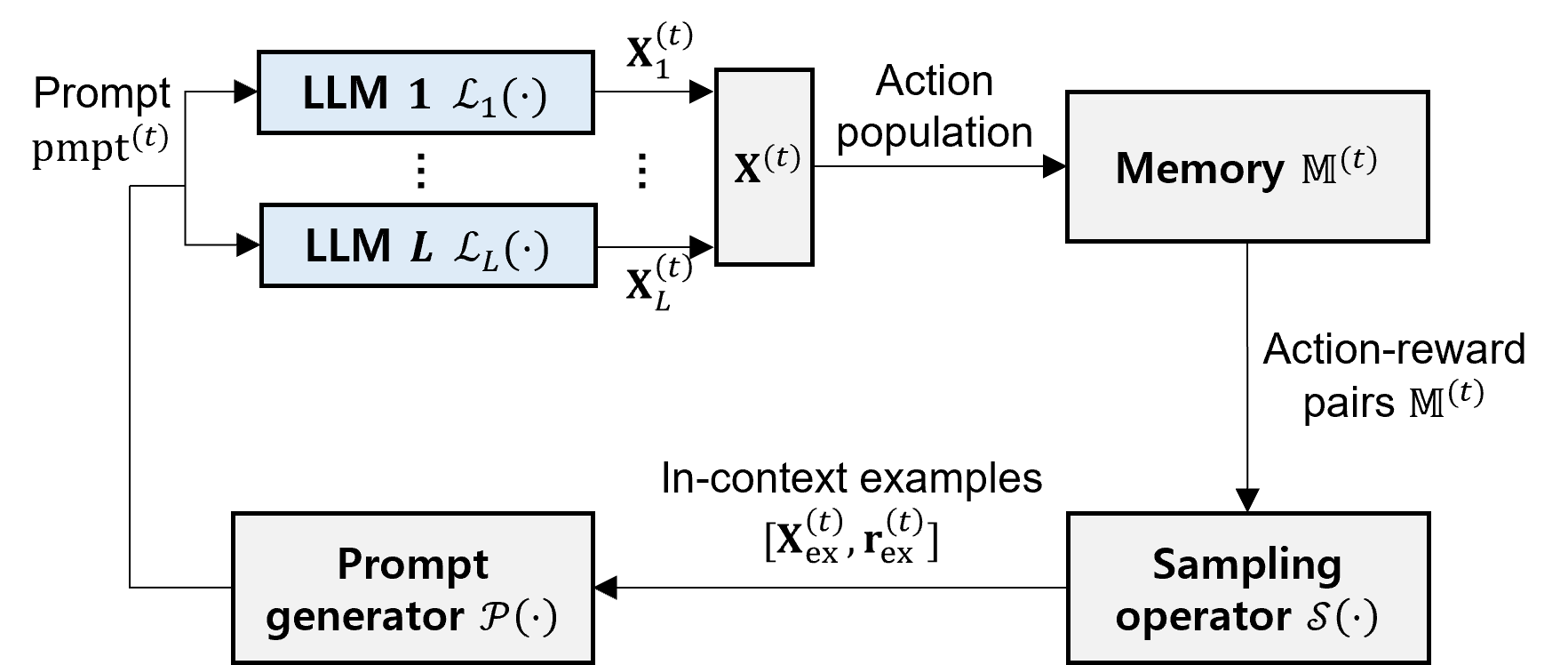}
    \caption{Multi-LLMO framework with $L$ LLMs \cite{HLee:24a}.}
    \label{fig:fig5m}
\end{figure}

Recently, this challenge has been addressed by using a multi-LLMO approach \cite{LEO,HLee:24a}. As illustrated in Fig. \ref{fig:fig5m}, this method employs $L$ LLMs to collaboratively solve BBO problems. The LLMs exchange their decisions through the shared memory $\mathbb{M}^{(t)}$. Hence, the in-context examples $[\mathbf{X}_{\sf{ex}}^{(t)},\mathbf{r}_{\sf{ex}}^{(t)}]$ contain successful decisions taken by a group of LLMs. Even though a certain LLM produces poor actions, the memory can also keep effective solutions generated by others. With the elitist sampling, the in-context examples accumulate the best action-reward pairs chosen by multiple LLMs. By providing such elitist solutions, we can mitigate the hallucination issue and improve the performance. It has been numerically shown that the convergence speed of the multi-LLMO is proportional to the number of LLMs $L$ \cite{HLee:24a}, while its theoretical analysis still remains unaddressed.

 
To unveil such a relationship, we analyze the convergence behavior of the multi-LLMO framework originally investigated in \cite{HLee:24a}. This architecture generalizes the single LLM counterpart \cite{OPRO}. As shown in Fig. \ref{fig:fig5m}, LLM $l$ ($l=1,\cdots,L$) produces candidate actions in parallel based on the prompt shown in Fig. \ref{fig:fig0}. Let $\mathcal{L}_{l}(\cdot)$ be LLM $l$ with the learned distribution $p_{\mathcal{L}_{l}}(\cdot|\cdot)$. At the $t$-th iteration, LLM $l$ determines its own action population $\mathbf{X}_{l}^{(t)}\in\mathbb{R}^{P\times D}$ based on $\sf{pmpt}^{(t-1)}$~as
\begin{align}
    \mathbf{X}^{(t)}_{l}=\mathcal{L}_{l}({\sf{pmpt}}^{(t-1)}),\ \text{for}\ l=1,\cdots,L. \label{eq:mLLM}
\end{align}
In practice, we can realize the multi-LLM inference \eqref{eq:mLLM} in parallel, thereby incurring no additional computational latency compared to the single LLM case. The output action populations of all LLMs $\mathbf{X}_{l}^{(t)}$, $\forall l$, are collected into the matrix $\mathbf{X}^{(t)}\triangleq[(\mathbf{X}_{1}^{(t)})^{T},\cdots,(\mathbf{X}_{L}^{(t)})^{T}]^{T}\in\mathbb{R}^{LP\times D}$, which contains $LP$ new actions. Similar to the single LLM case, we update the memory $\mathbb{M}^{(t)}$ as in \eqref{eq:memory}. This memory is shared across all LLMs for sampling the in-context example actions $\mathbf{X}_{\sf{ex}}^{(t)}$ in the subsequent iteration. To be specific, at the $t$-th iteration, the elitist sampling operator chooses the $P$ best actions among $P(L+1)$ actions of $\mathbf{X}^{(t-1)}$ and $\mathbf{X}_{\sf{ex}}^{(t-2)}$. 
In contrast, the LIFO sampling operator in the multi-LLM case can be set to sample the $P$ best actions only from the most recent decisions $\mathbf{X}^{(t-1)}$.

The multi-LLMO framework allows several LLMs to share their action-reward pairs through the memory unit $\mathbb{M}^{(t)}$. With the elitist sampling operator, we can choose the best actions among $L$ action populations $\mathbf{X}_{l}^{(t)}$, $\forall l$, which are individually produced by $L$ LLMs. Compared to the single LLM case, this results in more efficient in-context examples, which provide better information about the global optimal actions for multiple LLMs simultaneously. By doing so, we can leverage the collective intelligence of multiple LLMs.

Recent studies have numerically verified that the convergence speed of the multi-LLMO is proportional to the number of LLMs~\cite{HLee:24a}. However, the explicit impact of the number of LLMs on the convergence remains unexplored. In this section, we provide the theoretical analysis for this phenomenon by extending the results in Section \ref{sec:sec3}.

\subsection{Markov chain models}

We begin with the following lemma that establishes the Markov model of the multi-LLMO with the elitist sampling.

\begin{lemma}\label{prop:prop7}
With the elitist sampling, the multi-LLMO forms a Markov chain in \eqref{eq:markov} with the transition probability $q_{s\tilde{s}}=\Pr\{s^{(t)}=s|s^{(t-1)}=\tilde{s}\}$ given by
\begin{align}\label{eq:q_LLM}
    q_{s\tilde{s}}=\begin{cases}
        \prod\limits_{l=1}^{L}\!\sum\limits_{s^{\prime}\preceq  s}\lambda_{l,s^{\prime}\tilde{s}}- \prod\limits_{l=1}^{L}\!\sum\limits_{s^{\prime}\prec  s}\lambda_{l,s^{\prime}\tilde{s}}, & \text{for }s\succ \tilde{s}, \\
        \prod\limits_{l=1}^{L}\!\sum\limits_{s^{\prime}\preceq  \tilde{s}}\lambda_{l,s^{\prime}\tilde{s}}, &\text{for }s= \tilde{s}, \\
        0, &\text{for }s\prec \tilde{s},      
    \end{cases}
\end{align}
where 
$\lambda_{l,s\tilde{s}}\triangleq p_{\mathcal{L}_{l}}(\mathcal{T}_{\sf{BPE}}(\mathbf{X}^{(t)})=s|\mathcal{T}_{\sf{BPE}}(\mathbf{X}^{(t-1)}_{\sf{ex}})=\tilde{s},\mathcal{T}_{\sf{BPE}}(\mathbf{r}^{(t-1)}_{\sf{ex}}),\mathbf{z}_{\sf{fix}})$ is the transition probability of LLM $l$.
\end{lemma}
\begin{IEEEproof}
Let $s_{l}^{(t)}\triangleq\mathcal{T}_{\sf{BPE}}(\mathbf{X}_{l}^{(t)})\in\mathbb{S}$ be a state regarding $\mathbf{X}_{l}^{(t)}$ determined by LLM $l$. Since the elitist sampling produces the best $P$ actions, $s^{(t)}=\mathcal{T}_{\sf{BPE}}(\mathbf{X}_{\sf{ex}}^{(t)})$ can be expressed as 
\begin{align}
    s^{(t)}=\max\{s_{1}^{(t)},\cdots,s_{L}^{(t)},s^{(t-1)}\},
\end{align}
where the maximum operator defined for states $\max\{s,\tilde{s}\}$ produces $s$ if $s\succeq\tilde{s}$ and $\tilde{s}$ otherwise.
This implies that the multi-LLMO forms the Markov chain as $\cdots\rightarrow\mathbf{X}^{(t-1)}\rightarrow\mathbf{X}^{(t-1)}_{\sf{ex}}\rightarrow\mathbf{X}^{(t)}\rightarrow\mathbf{X}^{(t)}_{\sf{ex}}\rightarrow\cdots$, thereby leading to \eqref{eq:markov}. Next, we explain \eqref{eq:q_LLM} for three cases $s\prec \tilde{s}$, $s\succ \tilde{s}$, and $s= \tilde{s}$. First, it is straightforward to prove that $q_{s\tilde{s}}=0$ for $s\prec \tilde{s}$ since the elitist sampling is guaranteed to obtain improved action populations at each iteration, i.e., $s^{(t)}\succ  s^{(t-1)}$. 


Let us define the maximum operator over states $\max\{\tilde{s},s^{\prime}\}$ which produces $\tilde{s}$ if $\tilde{s}\succeq s^{\prime}$ and $s^{\prime}$ otherwise. Denoting $s_{\sf{max}}^{(t)}\triangleq\max_{l}s_{l}^{(t)}$, the transition probability $q_{s\tilde{s}}$ for $s\succ \tilde{s}$ can be calculated as
\begin{subequations}
\begin{align}
    q_{s\tilde{s}}&=\Pr\{\max\{s_{{\sf{max}}}^{(t)},s^{(t-1)}\}=s|s^{(t-1)}=\tilde{s}\}\\
    &=\Pr\{s_{{\sf{max}}}^{(t)}=s|s^{(t-1)}=\tilde{s}\}\\
    &=\Pr\{s_{\sf{max}}^{(t)}\preceq  s|s^{(t-1)}=\tilde{s}\} -\Pr\{s_{\sf{max}}^{(t)}\prec  s|s^{(t-1)}=\tilde{s}\}. \label{eq:qss}
\end{align}
\end{subequations}
Since $s_{l}^{(t)}$ for $l=1,\cdots,L$ are conditionally independent for a given $s^{(t-1)}$, it follows
\begin{subequations}
\begin{align}
    &\Pr\{s_{\sf{max}}^{(t)}\preceq  s|s^{(t-1)}=\tilde{s}\}\\
    &=\prod_{l=1}^{L}\Pr\{s_{l}^{(t)}\preceq  s|s^{(t-1)}=\tilde{s}\}=\prod_{l=1}^{L}\sum_{s^{\prime}\preceq s}\lambda_{l,s^{\prime}\tilde{s}}. \label{eq:ymax}
\end{align}
\end{subequations}
Similarly, we have $\Pr\{s_{\sf{max}}^{(t)}\prec  s|s^{(t-1)}=\tilde{s}\}=\prod_{l=1}^{L}\sum_{s^{\prime}\prec s}\lambda_{l,s^{\prime}\tilde{s}}$. Plugging these into \eqref{eq:qss} results in \eqref{eq:q_LLM} for $s\succ \tilde{s}$. Finally, to derive $q_{s\tilde{s}}$ for $s= \tilde{s}$, we use the fact $\sum_{s\in\mathbb{S}}q_{s\tilde{s}}=q_{\tilde{s}\tilde{s}}+\sum_{s\succ \tilde{s}}q_{s\tilde{s}}=1$, which leads to \eqref{eq:q_LLM}. This completes the proof.
\end{IEEEproof}

According to Lemma \ref{prop:prop7}, the transition matrix $\mathbf{Q}_{\sf{LLM}}$ of the multi-LLMO with the elitist sampling, which collects $q_{s\tilde{s}}$ in \eqref{eq:q_LLM} as its $(s,\tilde{s})$-th element, exhibits the same properties stated in Lemma \ref{prop:prop4}. Therefore, the optimality of the multi-LLMO is directly achieved by applying Theorem \ref{prop:prop5}. In contrast, with the LIFO sampling, the transition probability becomes $q_{s\tilde{s}}=\prod_{l=1}^{L}\!\sum_{s^{\prime}\preceq s}\!\lambda_{l,s^{\prime}\tilde{s}}- \prod_{l=1}^{L}\!\sum_{s^{\prime}\prec  s}\!\lambda_{l,s^{\prime}\tilde{s}}$ for any $s,\tilde{s}\in\mathbb{S}$, which is always positive. Combining this with Theorem \ref{prop:prop5}, we can conclude that the multi-LLMO with the LIFO sampling cannot identify the globally optimal solution.

\subsection{Convergence rate analysis}\label{sec:sec4B}
We analyze the convergence property of the multi-LLMO by examining the average convergence rate (ACR). The ACR at the $t$-th iteration, denoted by $\gamma^{(t)}$, is defined as \cite{ACR0,ACR1,ACR2,HJun:16}
\begin{align}
    \gamma^{(t)}=\left(\frac{|r^{\star}-\bar{r}^{(t)}|}{|r^{\star}-\bar{r}^{(0)}|}\right)^{1/t},\label{eq:logacr}
\end{align}
where $r^{\star}$ is the global optimal value of the problem in \eqref{eq:P}, $\bar{r}^{(t)}\triangleq\mathbb{E}_{s^{(t)}}[r^{(t)}_{\sf{best}}]$ equals the expected best reward $r^{(t)}_{\sf{best}}$ at the $t$-th iteration (See Algorithm \ref{alg:alg1}), and $|r^{\star}-\bar{r}^{(t)}|$ accounts for the optimality gap at the $t$-th iteration. A small ACR means that the expected reward $\bar{r}^{(t)}$ is close to the optimal reward value $r^{\star}$, demonstrating fast convergence. With the elitist sampling, the LLMO is guaranteed to have $|r^{\star}-\bar{r}^{(0)}|\geq |r^{\star}-\bar{r}^{(t)}|$, and thus we have $\gamma^{(t)}\in[0,1]$. 
In what follows, we provide the asymptotic behavior of the ACR.
\begin{lemma}\label{prop:prop8}
For any initial distribution with $\Pr\{s^{(0)}=s\}>0$, $\forall s\in\mathbb{S}$, it follows
\begin{align}\label{eq:acr}
    \lim_{t\rightarrow\infty}\gamma^{(t)}=q_{\sf{max}}\triangleq\max_{s\in\mathbb{S}^{\prime}}q_{ss}.
\end{align}
\end{lemma}
\begin{IEEEproof}
Similar to \eqref{eq:P_LLM}, the transition matrix $\mathbf{Q}_{\sf{LLM}}$ of the multi-LLMO, whose $(s,\tilde{s})$-th element $q_{s\tilde{s}}$ is given as \eqref{eq:q_LLM}, can be constructed as
\begin{align}\label{eq:Q_LLM}
    \mathbf{Q}_{\sf{LLM}}=\begin{bmatrix}
        \mathbf{Q}_{1} & \mathbf{Q}_{2} \\
        \mathbf{0}_{|\mathbb{S}^{\prime}|\times|\mathbb{S}^{\star}|} & \mathbf{Q}_{3}
    \end{bmatrix},
\end{align}
where $\mathbf{Q}_{1}\in\mathbb{R}^{|\mathbb{S}^{\star}|\times|\mathbb{S}^{\star}|}$ and $\mathbf{Q}_{3}\in\mathbb{R}^{|\mathbb{S}^{\prime}|\times|\mathbb{S}^{\prime}|}$ are upper triangular matrices and $\mathbf{Q}_{2}\in\mathbb{R}^{|\mathbb{S}^{\star}|\times|\mathbb{S}^{\prime}|}$.
From \cite[Theorem 1]{HJun:16}, we have $\lim_{t\rightarrow\infty}\gamma^{(t)}=\rho(\mathbf{Q}_{3})$ under the random initialization scheme $\Pr\{s^{(0)}=s\}>0$, $\forall s\in\mathbb{S}$, where $\rho(\mathbf{U})$ denotes the largest absolute eigenvalue of $\mathbf{U}$. Since $\mathbf{Q}_{3}$ is upper triangular, its eigenvalues are the diagonal elements $q_{ss}$ for $s\in\mathbb{S}^{\prime}$, which are nonnegative. This completes the proof.
\end{IEEEproof}

From Lemma \ref{prop:prop8}, we can see that for any initial action population $\mathbf{X}^{(0)}$, the ACR $\gamma^{(t)}$ converges to a finite value $q_{\sf{max}}$ as $t$ gets larger. This indicates that the optimality gap $|r^{\star}-\bar{r}^{(t)}|$ of the multi-LLMO must converge to a bounded value $q_{\sf{max}}$. Nevertheless, since this guarantee is asymptotic, the rate at which the multi-LLMO approaches the optimal action remains unclear. To address this issue, we introduce the result in \cite{HJun:16} which discusses the ACR for finite $t$.
\begin{lemma} \label{lem:lem4}
Let $\mathbf{v}\in\mathbb{R}^{|\mathbb{S}|}$ be the eigenvector of $\mathbf{Q}_{3}$ corresponding to the eigenvalue $q_{\sf{max}}$. The ACR becomes $\gamma^{(t)}=q_{\sf{max}}$ for any $t$ with the initial distribution given by
\begin{align}
    \Pr\{s^{(0)}=s\}=\frac{[\mathbf{v}]_{s}}{\sum_{\tilde{s}}[\mathbf{v}]_{\tilde{s}}}.\label{eq:p_init}
\end{align}
\end{lemma}
\begin{IEEEproof}
Please refer to \cite{HJun:16}.
\end{IEEEproof}
From Lemma \ref{lem:lem4}, we can conclude that it is possible to determine the initial distribution which achieves $\gamma^{(t)}=q_{\sf{max}}$ for any finite $t$. This allows us to characterize the convergence rate in representative scenarios as in the following.
\begin{theorem}\label{cor:cor1}
With \eqref{eq:p_init}, the optimality gap $|r^{\star}-\bar{r}^{(t)}|$ decreases by a factor of $q_{\sf{max}}$ at each iteration as
\begin{align} \label{eq:linconv}
    |r^{\star}-\bar{r}^{(t)}|=q_{{\sf{max}}}|r^{\star}-\bar{r}^{(t-1)}|,
\end{align}
where $q_{{\sf{max}}}<1$. In addition, when all LLMs are identical as $\mathcal{L}_{1}(\cdot)=\cdots=\mathcal{L}_{L}(\cdot)$, \eqref{eq:linconv} boils down to
\begin{align}
    |r^{\star}-\bar{r}^{(t)}|=\lambda^{L}|r^{\star}-\bar{r}^{(t-1)}| \label{eq:acr2}
\end{align}
for some scalar $\lambda<1$.
\end{theorem}
\begin{IEEEproof}
We first prove the property in \eqref{eq:linconv} and the fact $q_{\sf{max}}<1$. Combining $\gamma^{(t)}=q_{\sf{max}}$ and \eqref{eq:logacr}, we have
\begin{align}
    \frac{(\gamma^{(t)})^{t}}{(\gamma^{(t-1)})^{t-1}}=\frac{|r^{\star}-\bar{r}^{(t)}|}{|r^{\star}-\bar{r}^{(t-1)}|}=q_{\sf{max}}. 
\end{align}
From Lemma \ref{prop:prop4}, it is not difficult to show that the transition matrix  $\mathbf{Q}_{2}$ from the non-optimal to optimal states in \eqref{eq:Q_LLM} has at least one positive element in each column. For this reason, the diagonal element $q_{ss}$ of $\mathbf{Q}_{3}$ is always smaller than 1, meaning that $q_{\sf{max}}=\max_{s\in\mathbb{S}^{\prime}}q_{ss}<1$ This completes the proof for \eqref{eq:linconv}.

Next, to show \eqref{eq:acr2}, we consider a special case with all LLMs are identical. In this case, the transition probabilities $\lambda_{l,s\tilde{s}}$, $\forall l$, are the same as
\begin{align}
    \lambda_{s\tilde{s}}=\lambda_{1,s\tilde{s}}=\cdots=\lambda_{L,s\tilde{s}}.
\end{align}
In this case, the transition probability $q_{ss}$ for $s\in\mathbb{S}^{\prime}$ in \eqref{eq:q_LLM} reduces to $q_{ss}=(\sum_{s^{\prime}\preceq  s}\lambda_{s^{\prime}s})^{L}<1$.
As a result, $q_{\sf{max}}$ in \eqref{eq:acr} can be calculated as
\begin{align}
    q_{\sf{max}}=\max_{s\in\mathbb{S}^{\prime}}\bigg(\sum_{s^{\prime}\preceq  s}\lambda_{s^{\prime}s}\bigg)^{L}=\bigg(\max_{s\in\mathbb{S}^{\prime}}\sum_{s^{\prime}\preceq  s}\lambda_{s^{\prime}s}\bigg)^{L}.
\end{align}
By setting $\lambda=\max_{s\in\mathbb{S}^{\prime}}\sum_{s^{\prime}\preceq  s}\lambda_{s^{\prime}s}$, we have \eqref{eq:acr2}. This completes the proof.
\end{IEEEproof}

Theorem \ref{cor:cor1} quantifies the rate of convergence for the multi-LLMO. For a general case with heterogeneous LLMs, i.e., $\mathcal{L}_{i}(\cdot)\neq\mathcal{L}_{j}(\cdot)$, $\forall i\neq j$, \eqref{eq:linconv} indicates that the multi-LLMO exhibits a linear convergence property with the rate $q_{\sf{max}}<1$. More precisely, in the semi-logarithmic scale, the optimality gap $|r^{\star}-\bar{r}^{(t)}|$ linearly decreases with respect to the iteration $t$. Therefore, the multi-LLMO should converge to the global optimal point as $t$ increases.

In the special case where all LLMs are identical, \eqref{eq:acr2} reveals that the convergence rate $q_{\sf{max}}$ exponentially decreases with the number of LLMs $L$. Hence, at each iteration $t$, the optimality gap is reduced by $\lambda^{L}$. Since $\lambda<1$, increasing $L$ accelerates the convergence of the LLMO. Also, in the semi-logarithmic scale, the slope of $\gamma^{(t)}$ with respect to $L$ equals $\log_{10}\lambda$, indicating that the asymptotic ACR linearly decreases with the number of LLMs $L$. These validate the effectiveness of involving multiple LLMs in improving the convergence speed of the LLMO framework, which provides the mathematical background for existing numerical demonstrations \cite{HLee:24a}.

\section{Performance Evaluation}\label{sec:sec5}

This section verifies our analysis of the LLMO framework through numerical results. Unless stated otherwise, our simulations utilize GPT-3.5-Turbo implemented with OpenAI API. 
To check the universality of the LLMO approach, we consider three different network management tasks in IFCs, BCs, and massive MIMO systems. The population size is set to $P=5$ and the initial action population $\mathbf{X}^{(0)}$ is chosen uniformly.


\subsection{Transmit power control in interference channels}
We take into account the power control problems for IFC with $D=3$ transmitter-receiver pairs. Such a scenario can characterize various practical wireless networks such as multi-cell systems \cite{HLee:21,WMMSE}, device-to-device communications \cite{HLee:22,HLee:22-IoT}, and ad-hoc networks \cite{HLee:21-TWC}. In this system, we consider the energy efficiency (EE) and spectral efficiency (SE) reward functions defined as \cite{HLee:21}
\begin{align}
    r_{\sf{IFC}}^{\sf{EE}}(\mathbf{x})=\sum_{d=1}^{D}\frac{f_{d}(\mathbf{x})}{P_{\sf{fix}}+P_{\sf{tx}}x_{d}}\ \text{and}\ r_{\sf{IFC}}^{\sf{SE}}(\mathbf{x})\!=
    \!\sum_{d=1}^{D}f_{d}(\mathbf{x}), 
\end{align}
where the action vector $\mathbf{x}$ is defined as $\mathbf{x}\triangleq[x_1,\cdots,x_D]^{T}$ with $x_{d}\in[0,1]$ ($d=1,\cdots,D$) being the power allocation ratio at transmitter $d$, $f_{d}(\mathbf{x})\triangleq\log\left(1+\frac{P_{\sf{tx}}|h_{dd}|^2x_{d}}{1+P_{\sf{tx}}\sum_{d^{\prime}\neq d}|h_{d^{\prime}d}|^2x_{d^{\prime}}}\right)$ stands for the achievable rate at receiver $d$, $h_{d^{\prime}d}$ is the channel coefficient from transmitter $d^{\prime}$ to receiver $d$, and $P_{\sf{tx}}=10$ W and $P_{\sf{fix}}=1$ W respectively denote the transmit power budget and static power consumption at transmitters. We instruct the LLMs to generate integer actions between $0$ and $999$ and then normalize outputs by $999$ to retrieve feasible actions $x_{d}\in[0,1]$ with fixed precision $N_{\sf{digit}}=3$.\footnote{One can instruct the LLM to directly generate floating-point actions $x_{d}\in[0,1]$ with three decimal places. This requires $N_{\sf{token}}=4$ tokens for each $x_{d}$. In contrast, an integer action can be encoded via only two tokens, thereby reducing the LLM inference complexity.} In all simulations, we plot the average reward performance over the Rayleigh fading channel samples $h_{d^{\prime}d}\sim\mathcal{CN}(0,1)$. For GPT-3.5-Turbo, the LLMO needs to process only $193$ tokens per iteration.


We consider the following optimization approaches.
\begin{itemize}
    \item \textit{Graph neural network (GNN) \cite{GNN}:} The learning-to-optimize approach \cite{HLee:19JSAC} is adopted with the GNN model. A two-layer GNN is constructed where each component neural network is realized with a two-layer multilayer perception with the leaky rectifier linear unit activation. The hidden and embedding dimensions are set to 256 and 32, respectively. The Adam optimizer with the learning rate $10^{-4}$ is used to train the GNN over $10^5$ epochs. At each epoch, randomly generated $5000$ Rayleigh fading samples are employed as training datasets.
    \item \textit{Local optimal:} We employ the fractional programming algorithm \cite{FP} for the EE and weighted minimum mean-squared error (WMMSE) \cite{WMMSE} for the SE, respectively. We depict the best reward over 50 random initial points.
    \item \textit{LLMO with elitist sampling (LLMO-E)}: The LLMO is implemented with the elitist sampling. 
    \item \textit{LLMO with LIFO sampling (LLMO-L)}: We adopt the LIFO sampling operator for the LLMO.
    \item \textit{GA \cite{GA}:} We utilize the GA with population size $P$ and the uniform crossover strategy with probability $0.5$. The parent portion is set to $0.3$. Also, the random mutation technique is employed where the mutation probability is fine-tuned for each problem. We execute $L$ independent trials and evaluate the best performance.
    \item \textit{BO \cite{BO}:} This probabilistic BBO algorithm leverages machine learning–based surrogate models to tackle unknown reward functions. We employ a random forest (RF) to represent the expected improved (EI) acquisition function. At each iteration, we sample the best $P$ solutions that maximize the EI, and then compute their reward values. These new samples are used to train the RF for the subsequent iterations. Similar to the GA, $L$ independent executions of the BO are carried out.
    \item \textit{Brute-force:} We uniformly choose $LP$ random actions at each iteration.
\end{itemize}
We employ the same number of iterations for the LLMO, GA, BO, and brute-force methods. By doing so, we can check the performance of diverse BBO schemes under the same number of reward evaluations \cite{LMEA}. Since the GNN and local optimal algorithms resort to mathematical models of reward functions, these provide ideal performance to the LLMO framework. 
For a fair comparison, hyperparameters of all benchmark methods are optimized for each application scenario.

\begin{figure}
\centering
    \subfigure[Average EE]{
        \includegraphics[width=.7\linewidth]{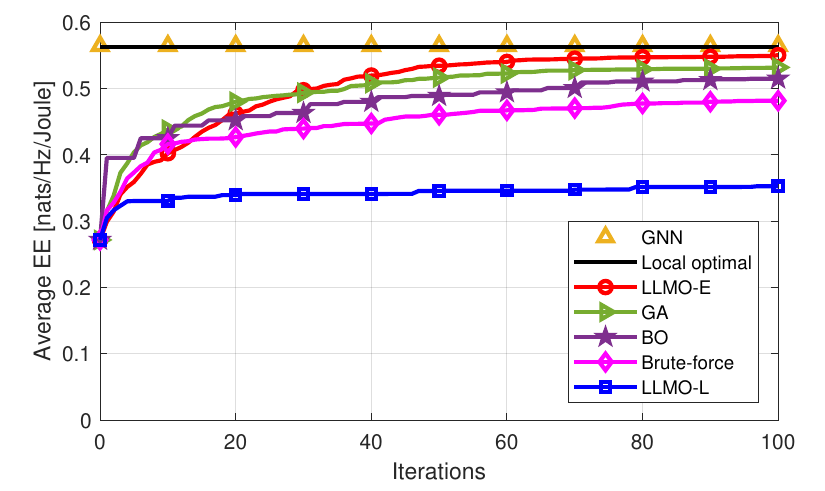}\label{fig:fig1a}
    }
    \subfigure[Average SE]{
        \includegraphics[width=.7\linewidth]{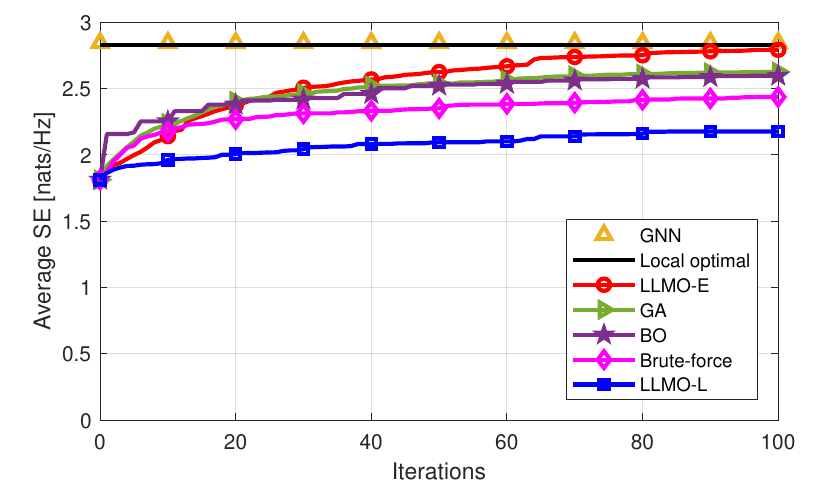}\label{fig:fig1b}
    }
    \caption{Convergence behavior of various schemes for $L=1$.}
    \label{fig:fig1}
\end{figure}

Fig. \ref{fig:fig1} presents the convergence behavior of various optimization algorithms with $L=1$ by evaluating the average EE and average SE in terms of the iterations. Regardless of the reward functions, the LLMO-E outperforms other BBO schemes, e.g., the GA and the BO. Since these methods require a careful optimization of various hyperparameters, they need to be fine-tuned for each given network scenario. In contrast, the LLMO does not need any human intervention, and thus it is suitable for a universal network optimizer. Even if LLMO-E operates without any mathematical models and hyperparameter tuning, it achieves almost identical performance to the GNN and the local optimal algorithms that heavily depend on models and handcraft designs. These results validate the theoretical analysis in Theorem \ref{prop:prop5} that proves the optimality of the LLMO-E for any given reward function. In addition, the LLMO-L fails to identify efficient actions for both the EE and SE maximization tasks. In fact, its performance is even lower than that of the brute-force method. This highlights the importance of the elitist sampling in designing the LLMO.

\begin{figure}
\centering
\includegraphics[width=.7\linewidth]{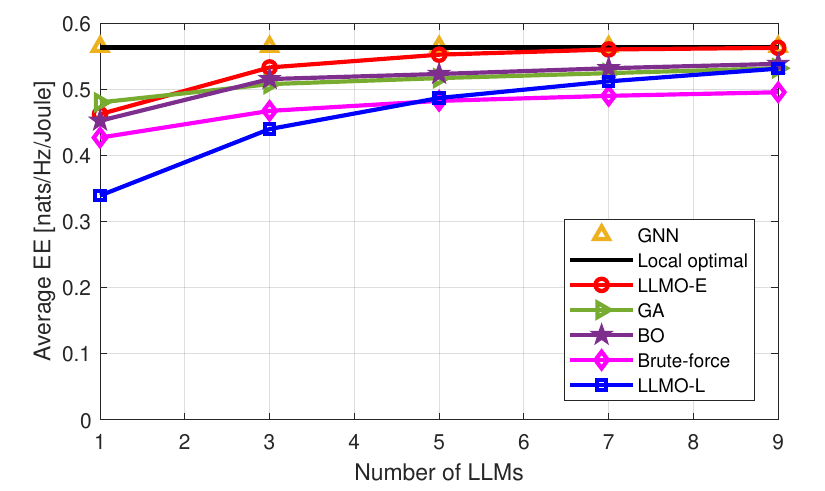}
\caption{Average EE with respect to $L$ with $t=20$.}
\label{fig:fig2}
\end{figure}

Fig. \ref{fig:fig2} plots the average EE performance at the $20$-th iteration with respect to $L$. The performance of the LLMO-E improves by employing more LLMs, validating the result in \eqref{eq:acr2}. With $L=5$ LLMs, the LLMO-E achieves $98\ \%$ of the local optimal performance and performs better than other methods. 
This implies the effectiveness of the multi-LLMO with the elitist sampling operator.

\begin{figure*}
\centering
    \subfigure[$L=1$]{
        \includegraphics[width=.34\linewidth]{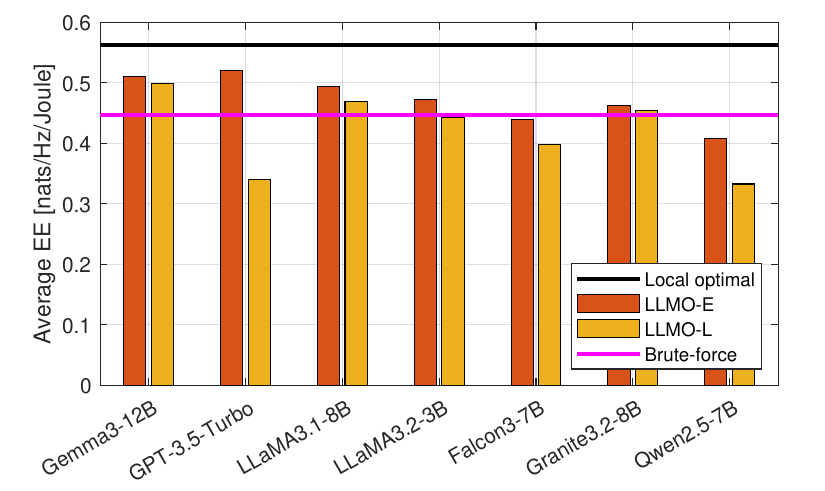}\label{fig:fig10a}
    }\hspace{-9mm}
    \subfigure[$L=3$]{
        \includegraphics[width=.34\linewidth]{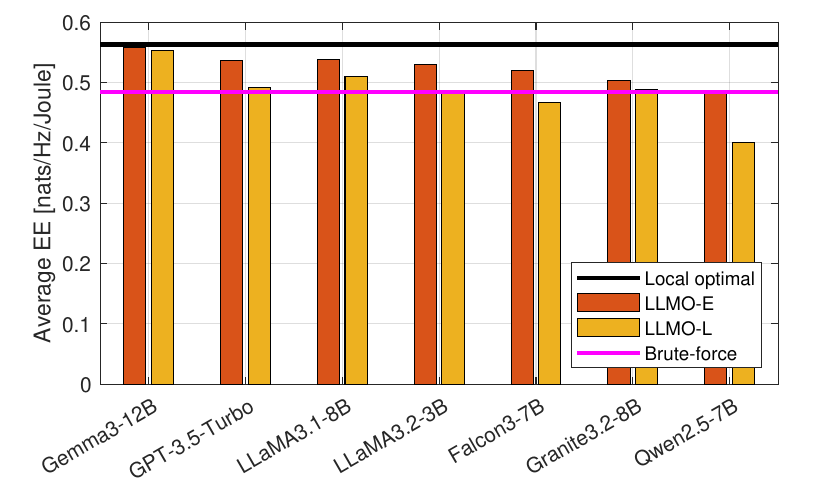}\label{fig:fig10b}
    }\hspace{-9mm}
    \subfigure[$L=5$]{
        \includegraphics[width=.34\linewidth]{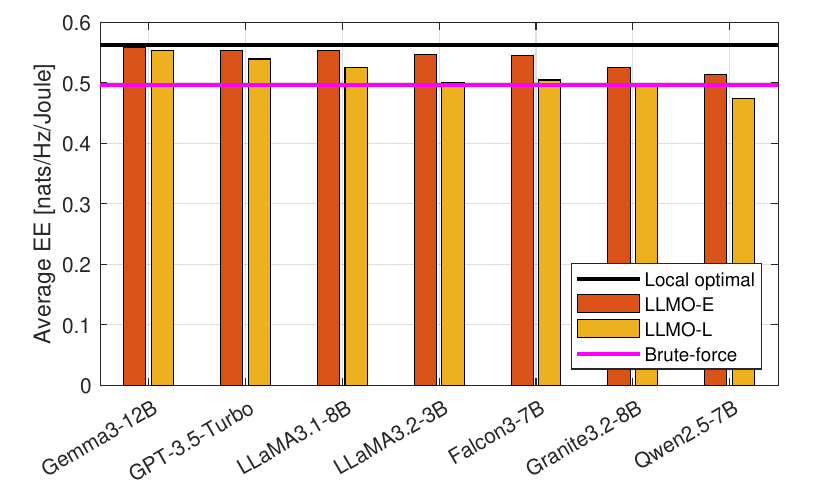}\label{fig:fig10c}
    }
    \caption{Average EE performance for various LLMO configurations with $t=40$.}
    \label{fig:fig10}
\end{figure*}

Fig. \ref{fig:fig10} presents the average EE performance of the LLMO at $t=40$ for different numbers of LLMs $L$. We examine diverse open source models including Gemma3-12B, LLaMA3.1-8B, LLaMA3.2-3B, Falcon3-7B, and Qwen2.5-7B. We can see that the LLMO-E outperforms the LLMO-L regardless of the language models. The average EE of the LLMO-E gradually increases with the number of LLMs $L$ and approaches the locally optimal performance. This validates that our analysis in \eqref{eq:acr2} holds for any given language model. For $L=1$, GPT-3.5-Turbo performs better than others, whereas Gemma3-12B exhibits the best performance in the multi-LLM cases. The smallest model, i.e., LLaMA3.2-3B, shows comparable EE performance to that of larger models such as LLaMA3.1-8B. Thus, it can be concluded that lightweight LLMs would be a good choice for implementing the LLMO in practice.

\begin{figure}
    \centering
    \includegraphics[width=.7\linewidth]{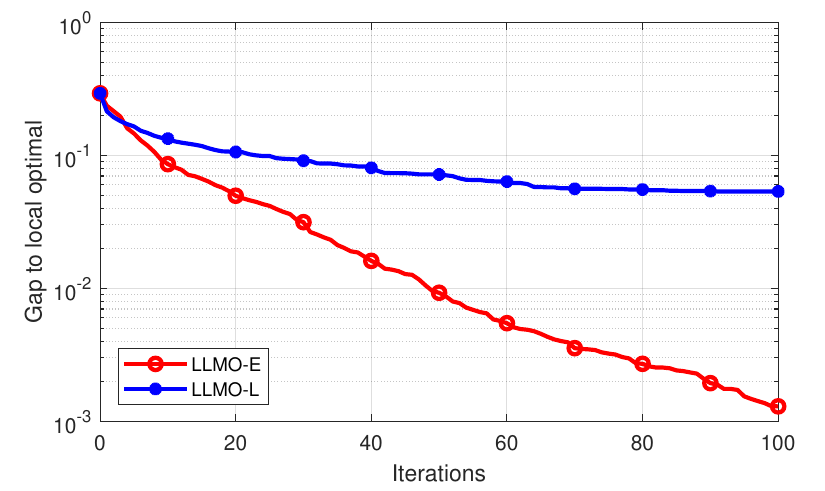}
    \caption{Convergence behavior of LLMO with heterogeneous LLMs.}
    \label{fig:figR4C6}
\end{figure}

Next, we present numerical results to verify our convergence rate analysis in Theorem \ref{cor:cor1}. First, Fig. \ref{fig:figR4C6} demonstrates the linear convergence property in \eqref{eq:linconv} by plotting the reward gap of the LLMO to the locally optimal algorithm \cite{PF} in the EE maximization task. Here, we consider a generic heterogeneous case where the LLMO is realized with three different models, i.e., GPT-3.5-Turbo, LLaMA3.1-8B, and Falcon3-7B. For the LLMO-E, the gap linearly decays in the semi-logarithmic scale, which confirms the theoretical convergence rate analysis in \eqref{eq:linconv}. In contrast, the LLMO-L converges slowly and even becomes saturated after the 80-th iteration. This verifies the impact of the elitist sampling on the convergence rate.

\begin{figure}
    \centering
    \includegraphics[width=.7\linewidth]{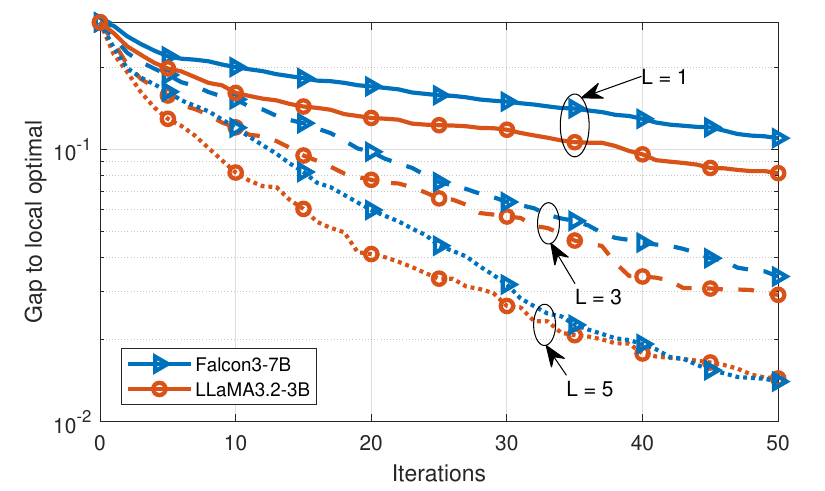}
    \caption{Convergence behavior of LLMO-E with various $L$.}
    \label{fig:figR1C4}
\end{figure}

Fig. \ref{fig:figR1C4} presents the convergence behavior of the LLMO-E for the EE maximization problem with various $L$. To validate \eqref{eq:acr2}, the multi-LLMO is realized with the identical LLMs including LLaMA3.2-3B and Falcon3-7B. For all simulated $L$ and language models, the gap decays linearly in the semi-logarithmic scale, confirming the linear convergence rate in \eqref{eq:acr2}. As $L$ increases, the curves become steeper, which demonstrates the acceleration brought by the multi-LLMO. Our analysis \eqref{eq:acr2} reveals that the slope of the optimality gap is given by $L\log_{10}\lambda$. Using the curve-fitting techniques, the slope for Falcon3-7B with $L=1$ equals $\log_{10}\lambda=5.7\times10^{-3}$. According to \eqref{eq:acr2}, the slopes for $L=3$ and $L=5$ are predicted as $1.7\times10^{-2}$ and $2.8\times10^{-2}$, respectively. Through the simulations, these are respectively computed as $1.8\times10^{-2}$ and $2.7\times10^{-2}$, which are almost identical to the theoretical analysis. A similar observation can be obtained for LLaMA3.2-3B. These results prove the correctness of our analysis in Theorem \ref{cor:cor1}.

\begin{figure}
\centering
\includegraphics[width=.7\linewidth]{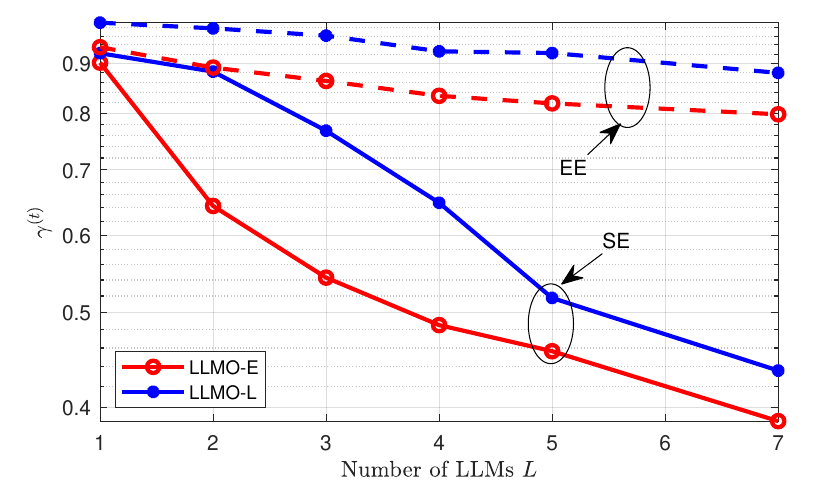}
\caption{Average convergence rate with respect to $L$ with $t=50$.}
\label{fig:fig3}
\end{figure}

Our analysis in \eqref{eq:acr2} reveals that the slope of the ACR $\gamma^{(t)}$ in terms of the number of LLMs $L$ becomes a negative number in the semi-logarithmic scale. To verify this, Fig. \ref{fig:fig3} depicts a semi-logarithmic plot of $\gamma^{(t)}$ at $t=50$ as a function of $L$ for the EE and SE maximization problems with GPT-3.5-Turbo. Here, $r^{\star}$ in \eqref{eq:logacr} is replaced with the best local optimal performance. For both EE and SE cases, $\gamma^{(t)}$ of the LLMO-E and LLMO-L linearly decreases with $L$, which confirms our analysis in \eqref{eq:acr2}. The LLMO-E shows a smaller $\gamma^{(t)}$ than that of the LLMO-L. Thus, we can conclude that the benefit of using multiple LLMs becomes more pronounced with the elitist sampling.

\subsection{Transmit power control in multi-user broadcast channels}

Next, we consider the transmit power allocation in the BC with $D=3$ users. In this case, we delve into the ability of the LLMO to address the sum-power constraint. The SE reward function of the multi-user BC system is given by
\begin{align}
    r_{\sf{BC}}^{\sf{SE}}(\mathbf{x})=\sum_{d=1}^{D}\log\left(1+\frac{P_{\sf{tx}}|h_{d}|^2x_{d}}{1+P_{\sf{tx}}|h_{d}|^2\sum_{d^{\prime}\neq d}x_{d^{\prime}}}\right), \label{eq:se_bc}
\end{align}
where $h_{d}$ is the channel coefficient for user $d$ and $x_{d}\in[0,1]$ accounts for the power allocation ratio for user $d$ subject to the sum-power constraint $\sum_{d=1}^{D}x_{d}\leq1$. The locally optimal performance for this task is obtained using the WMMSE algorithm \cite{WMMSE}, where the best performance over $50$ uniformly random initial points is plotted. We depict the average SE performance over the Rayleigh fading channels $h_{d}\sim\mathcal{CN}(0,1)$. We adopt the following constraint handling approaches.
\begin{itemize}
    \item \textit{Language constraint:} In the task description part of the input prompt in Fig. \ref{fig:fig1}, we add the context information of the sum power constraint as ``The action vector should satisfy 0$<$=x\_d$<$=1 for d=1, ..., D and $\backslash$sum\_\{d=1\}\^{}\{D\}x\_d$<$=1.''
    \item \textit{Penalty method:} Instead of informing the sum-power constraint, a very small reward value, e.g., $-1000$, is provided if LLMs produce infeasible actions.
\end{itemize}
The language constraint approach efficiently exploits the reasoning ability of LLMs by including closed-form constraint expressions directly in the input prompt. However, it cannot handle black-box constraints with no mathematical expressions. In such cases, we can employ the penalty method, which only requires binary information about constraint violations.

\begin{figure}
\centering
\includegraphics[width=.7\linewidth]{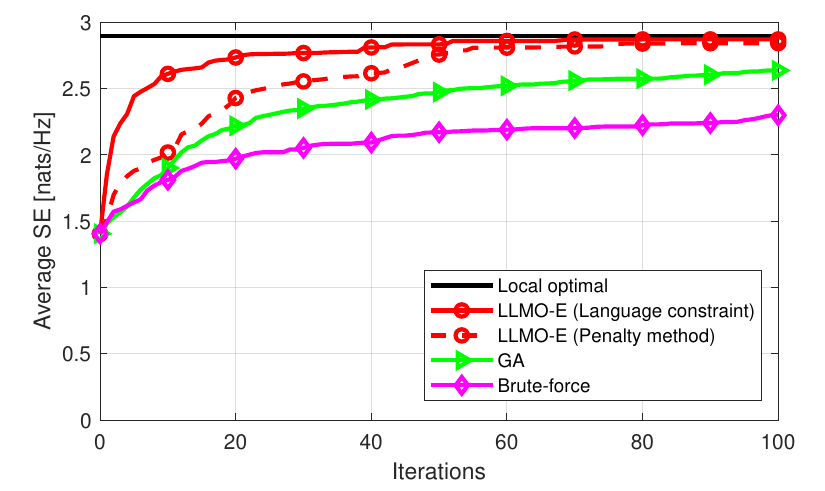}
\caption{Convergence behavior of various schemes with $L=3$.}
\label{fig:fig5}
\end{figure}

Fig. \ref{fig:fig5} shows the average SE performance with respect to the iterations for $L=3$ and $P_{\sf{tx}}=10$. The LLMO-E with the language constraint outperforms all other schemes and achieves the local optimal performance. The penalty method exhibits a slow convergence behavior compared to the language constraint method, but its performance approaches that of the WMMSE algorithm after the $60$-th iteration. Thus, we can conclude that the LLMO can handle black-box constraint functions through a simple penalty mechanism.

\begin{figure}
\centering
\includegraphics[width=.7\linewidth]{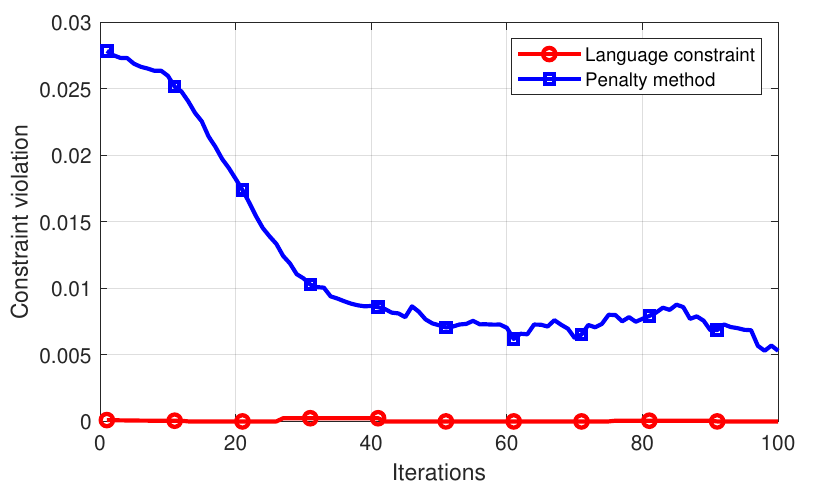}
\caption{Constraint violation of LLMO-E with $L=3$.}
\label{fig:fig_R4C8}
\end{figure}

To assess the constraint handling methods, Fig. \ref{fig:fig_R4C8} presents the constraint violation $\max\{0,\sum_{d=1}^{D}x_{d}-1\}$ for the sum-power constraint $\sum_{d=1}^{D}x_{d}\leq1$ with respect to iterations for the LLMO-E with $L=3$. The constraint violation lies in $[0,1]$, where small values indicate that actions generated by the LLMO are closer to the feasible region. 
The constraint violation of the language constraint method is zero almost everywhere, indicating that the LLMO-E generates feasible power control actions. In contrast, the penalty method begins with a large violation, which drops below $10^{-2}$ after the $30$-th iteration. As a consequence, as observed from Fig. \ref{fig:fig5}, the average SE of the penalty method improves more slowly than that of the language constraint method, and it eventually approaches the local optimum. This demonstrates that the penalty reward successfully guides the LLMO-E toward a feasible action space at convergence.

\begin{figure}
\centering
\includegraphics[width=.7\linewidth]{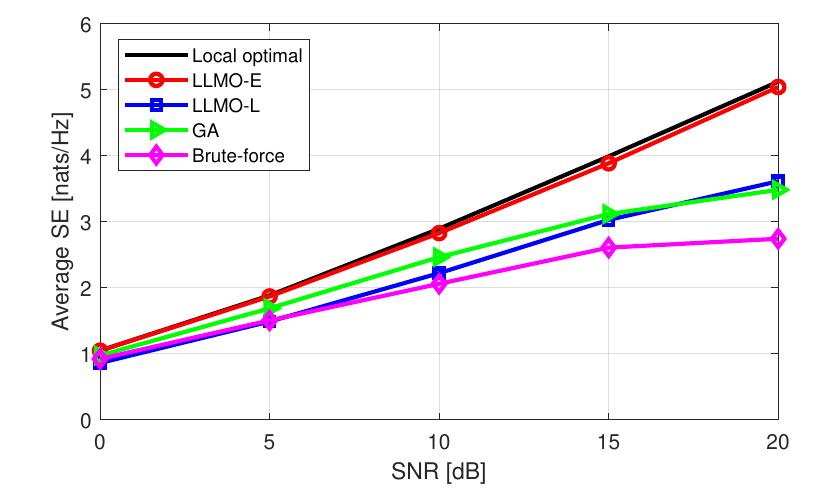}
\caption{Average SE with respect to SNR with $L=3$ and $t=100$.}
\label{fig:fig7}
\end{figure}

Fig. \ref{fig:fig7} illustrates the average SE performance by varying the signal-to-noise ratio (SNR) $P_{\sf{tx}}$ for $L=3$ and $t=100$. We employ the language constraint method both for the LLMO-E and LLMO-L. The LLMO-E achieves the local optimal performance for all simulated SNR. In contrast, the performance of other baseline methods, in particular, the GA, is degraded as the SNR grows. A large $P_{\sf{tx}}$ results in a significant fluctuation in the SE reward function \eqref{eq:se_bc}. This typically leads to premature convergence, necessitating an appropriate reward scaling strategy, which in turn requires additional hyperparameter tuning \cite{Vladik:93}. The LLMO-E can overcome this challenge without invoking any fine-tuning processes. 

\subsection{Massive MIMO systems}
We consider the cell-average EE maximization problem in multi-user massive MIMO systems whose reward function has no analytical expressions. Our goal is to optimize the number of base station (BS) antennas $M$, the number of user equipments (UEs) $K$, and the downlink transmit power at the BS $p_{\sf{dl}}$, collectively forming a three-dimensional action vector $\mathbf{x}=[M,K,p_{\sf{dl}}]^{T}$. Such a network planning problem is essential to realize green wireless systems \cite{BEmil:15}. To validate the LLMO in a real-world propagation environment, we adopt the DeepMIMO dataset \cite{DeepMIMO}. We consider the ASU campus scenario comprising the BS equipped with a uniform linear array and randomly deployed single antenna UEs. This digital twin platform generates channel realizations via ray tracing engines that capture realistic features such as reflection, diffraction, and diffuse scattering.

To estimate the channels, the UEs transmit orthogonal pilot sequences of length $K$ with the transmit power of $p_{\sf{ul}}=23\ \text{dBm}$. Let $\mathbf{h}_{k}\in\mathbb{C}^{M}$ be the channel vector from the BS to UE $k$. Also, we denote $\hat{\mathbf{h}}_{k}\in\mathbb{C}^{M}$ as the estimated channel vector obtained using the linear minimum mean-squared-error estimator. The BS employs the maximum ratio transmission beamforming. With equal power allocation, the achievable rate of UE $k$ is defined as \cite{BEmil:15}
\begin{align}
    f_{k}(\mathbf{x})\!=\!\left(1-\frac{K}{C}\right)\!\log\left(1\!+\!\frac{\frac{p_{\sf{dl}}}{K}|\mathbf{h}_{k}^{H}\hat{\mathbf{h}}_{k}|^{2}/||\hat{\mathbf{h}}_{k}||^2}{\sigma^{2}\!+\!\frac{p_{\sf{dl}}}{K}\sum_{j\neq k}|\mathbf{h}_{k}^{H}\hat{\mathbf{h}}_{j}|^{2}/||\hat{\mathbf{h}}_{j}||^2}\right),
\end{align}
where $\sigma^{2}=-96\ \text{dBm}$ is the noise power and $C=1800$ represents the number of symbols in a coherence block.

The cell-average EE reward function is computed as
\begin{align}
    r_{\sf{mMIMO}}^{\sf{EE}}(\mathbf{x})=\mathbb{E}_{\mathbf{H},\hat{\mathbf{H}}}\left[\frac{\sum_{k=1}^{K}f_{k}(\mathbf{x})}{g(\mathbf{x})}\right],\label{eq:EE_mMIMO}
\end{align}
where $g(\mathbf{x})$ stands for the network power consumption~\cite{BEmil:15} which depends on the numbers of antennas $M$ and users $K$, BS transmit power $p_{\sf{dl}}$, and data rate $f_{k}(\mathbf{x})$. Here, the average EE reward is considered where the expectation is taken over the joint distribution of the actual channel $\mathbf{H}\triangleq[\mathbf{h}_{1},\cdots,\mathbf{h}_{K}]$ and its estimate $\hat{\mathbf{H}}\triangleq[\hat{\mathbf{h}}_{1},\cdots,\hat{\mathbf{h}}_{K}]$. Due to the complicated formulas and the ray tracing channels, no closed-form expressions are available for $r_{\sf{mMIMO}}^{\sf{EE}}(\mathbf{x})$.
For this BBO task, we evaluate the associated reward $r_{\sf{mMIMO}}^{\sf{EE}}(\mathbf{x})$ through the digital twin-based simulation for each action $\mathbf{x}$ taken by LLMs. 
The feasible space is set to $M\leq256$, $K\leq 256$, and $p_{\sf{dl}}\leq50\ \text{dBm}$. The global optimal action has been calculated using exhaustive search as $\mathbf{x}^{\star}=[77,72,39\ \text{dBm}]^{T}$. 

\begin{figure}
\centering
\includegraphics[width=.7\linewidth]{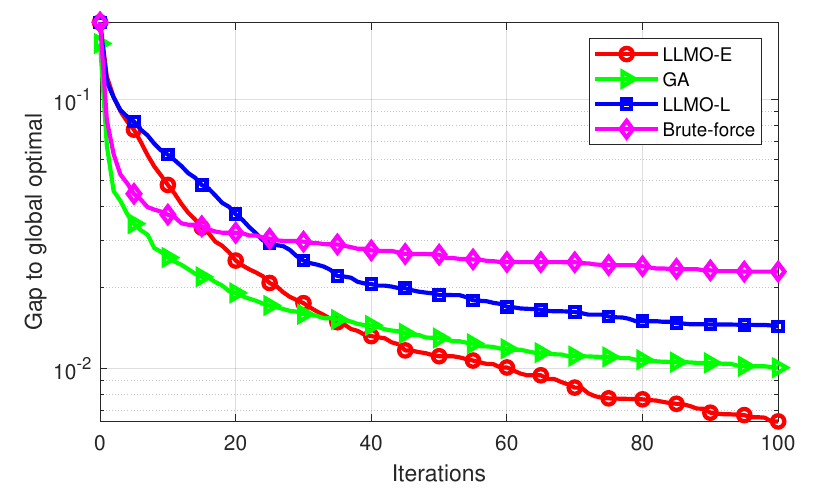}
\caption{Convergence behavior for cell-average EE reward with $L=3$.}
\label{fig:fig8}
\end{figure}

Fig. \ref{fig:fig8} illustrates the optimality gap of various schemes with $L=3$. We run $50$ independent executions of each method and plot the average performance. In the beginning, the LLMO-E performs worse than the GA and brute-force benchmarks. However, its optimality gap quickly decreases after the $30$-th iteration and achieves the best performance at the $100$-th iteration. In contrast, the LLMO-L exhibits premature behaviors and fails to identify the optimal action. This result demonstrates the viability of the LLMO-E in tackling BBO problems where closed-form expressions and gradients of the reward function in \eqref{eq:EE_mMIMO} are unavailable.

\section{Concluding Remarks and Future Works}\label{sec:sec6}
This paper has established a theoretical foundation for the LLMO in tackling black-box network management tasks. The overall procedure can be modeled as a finite-state Markov chain, which generates enhanced actions based on the past decision. The LLMO realized with the elitist sampling strategy is guaranteed to converge to the global optimum. In addition, we have presented a mathematical analysis of the convergence speed, highlighting the effectiveness of using multiple LLMs. The proposed theoretical results have been further validated through extensive simulations on various network management problems.
Simulation results have confirmed the superiority of the LLMO over traditional BBO algorithms.

Since the LLMO depends on a series of LLM inferences, its computational cost would be a critical issue for real-world networks. One viable approach is to adopt a suitable early stopping technique. By doing so, the number of LLM inferences can be reduced without the performance degradation.

In large-scale networks, the LLMO must process numerous tokens, leading to performance degradation \cite{LEO}. To address this scalability challenge, we can decouple action vectors into multiple blocks of variables \cite{HLee:24b}. 
Consequently, the LLMO can scale to handle large-scale problems.

Current LLM training datasets lack advanced wireless communication texts, which introduce internal biases. We can leverage telecom-specific LLMs \cite{TelecomGPT}. Such models can grasp the context of wireless networks, thereby mitigating biases and enhancing optimization performance.

The performance of the LLMO could be further enhanced by carefully designing prompts. It is thus necessary to find the optimal prompt templates in terms of the optimization capability. An LLM-based prompt optimization framework \cite{OPRO,ANie:23} can be adopted.

Exploring suitable applications for the LLMO framework is worth pursuing. Complicated network management tasks such as network slicing problems \cite{JTong:24} are potential candidates. The LLMO combined with digital twins \cite{KQiu:24} can generate efficient solutions to real-world scenarios.

Our theoretical results remain valid for nonconvex network management tasks. Nevertheless, as highlighted in \cite{LEO,BHuang:24,HLee:24a}, the current LLMO framework might struggle when the reward contains many local optima. A rigorous analysis of the LLMO under highly nonconvex reward functions is an important direction for future work.

For practical implementation, a decentralized LLMO needs to be further investigated. Although this has been recently studied in \cite{HLee:24b}, its theoretical analysis has not yet been investigated.

\begin{appendices}
\section{Proof of Lemma \ref{prop:prop1}}\label{app:appA}
\begin{figure}
    \centering
    \includegraphics[width=.7\linewidth]{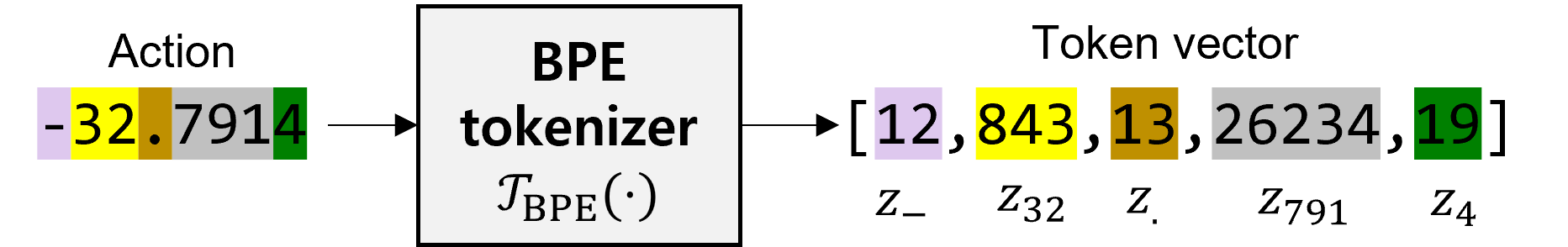}
    \caption{An example of BPE tokenizer}
    \label{fig:fig6m}
\end{figure}

We will first show that the vocabulary $\mathbb{T}$ of the LLMO is given by \eqref{eq:Z}. The BPE tokenizer groups up to three consecutive digits before or after the decimal point into a single token. Any remaining digits are then converted into separate tokens. It also treats the minus sign “-” and decimal point “.” as independent tokens. An example of such a tokenization procedure is presented in Fig. \ref{fig:fig6m} where an input action “$-32.7914$” is transformed into a token vector $[z_{\sf{-}},z_{\sf{32}},z_{\sf{.}},z_{\sf{791}},z_{\sf{4}}]=[12,843,13,26234,19]$.

Since we instruct the LLM to produce the new action population matrix $\mathbf{X}$ in CSV format, each element of $\mathbf{X}$ ends with commas “,” or newline characters “$\backslash$n”. Therefore, $\mathbb{T}$ includes tokens associated with one-digit numbers $(0,1,\cdots,9)$, two-digit numbers $(00,01,\cdots,99)$, three-digit numbers $(000,001,\cdots,999)$, the decimal point, minus sign, comma, and newline character, thereby leading to \eqref{eq:Z}. Thus, a fixed-precision floating-point number with $N_{\sf{digit}}$ digits requires $\lceil N_{\sf{digit}}/3\rceil$ tokens. We need three extra tokens for the decimal point, minus sign, comma, and newline character. Consequently, the number of tokens required for representing each element of $\mathbf{X}$ becomes $N_{\sf{token}}=\lceil N_{\sf{digit}}/3\rceil+3$. Since $\mathbf{X}$ contains $PD$ elements, its token vector $\mathcal{T}_{\sf{BPE}}(\mathbf{X})$ lies in a finite space $\mathbb{S}=\mathbb{T}^{PDN_{\sf{token}}}$. This completes the proof.



\section{Proof of Theorem \ref{prop:prop3}} \label{app:appC}

Let $\mathbf{X}_{\sf{ex}}^{(1:t-1)}\triangleq\{\mathbf{X}_{\sf{ex}}^{(\tau)}:\forall \tau=0,1,\cdots,t-1\}$ be the set of past in-context examples. To prove Theorem \ref{prop:prop3}, it suffices to show that the current in-context examples $\mathbf{X}_{\sf{ex}}^{(t)}$ depend only on $\mathbf{X}_{\sf{ex}}^{(t-1)}$ but not its entire history $\mathbf{X}_{\sf{ex}}^{(1:t-1)}$ as
\begin{align}
    \Pr\{\mathbf{X}_{\sf{ex}}^{(t)}|\mathbf{X}_{\sf{ex}}^{(1:t-1)}\}=\Pr\{\mathbf{X}_{\sf{ex}}^{(t)}|\mathbf{X}_{\sf{ex}}^{(t-1)}\}. \label{eq:pXex}
\end{align}
We begin by expanding $\Pr\{\mathbf{X}_{\sf{ex}}^{(t)}|\mathbf{X}_{\sf{ex}}^{(1:t-1)}\}$ as
\begin{subequations}\label{eq:pspl}
\begin{align}
    &\Pr\{\mathbf{X}_{\sf{ex}}^{(t)}|\mathbf{X}_{\sf{ex}}^{(1:t-1)}\}=\!\!\!\!\!\!\sum_{\mathcal{T}_{\sf{BPE}}(\mathbf{X}^{(t)})\in\mathbb{S}}\!\!\!\!\!\!\!\!\!\Pr\{\mathbf{X}^{(t)}_{\sf{ex}},\mathbf{X}^{(t)}|\mathbf{X}^{(1:t-1)}_{\sf{ex}}\}\\
    &=\!\!\!\!\!\!\sum_{\mathcal{T}_{\sf{BPE}}(\mathbf{X}^{(t)})\in\mathbb{S}}\!\!\!\!\!\!\!\!\!\Pr\{\mathbf{X}^{(t)}_{\sf{ex}}|\mathbf{X}^{(t)},\mathbf{X}^{(1:t-1)}_{\sf{ex}}\}\Pr\{\mathbf{X}^{(t)}|\mathbf{X}^{(1:t-1)}_{\sf{ex}}\}, \label{eq:psplc}
\end{align}
\end{subequations}
where the summations are taken over all possible tokens $\mathcal{T}_{\sf{BPE}}(\mathbf{X}^{(t)})\in\mathbb{S}$. In what follows, we show that two conditional distributions $\Pr\{\mathbf{X}^{(t)}_{\sf{ex}}|\mathbf{X}^{(t)},\mathbf{X}_{\sf{ex}}^{(1:t-1)}\}$ and $\Pr\{\mathbf{X}^{(t)}|\mathbf{X}_{\sf{ex}}^{(1:t-1)}\}$ in \eqref{eq:psplc} can be written as
\begin{subequations}\label{eq:p_ind}
\begin{align}
    &\Pr\{\mathbf{X}_{\sf{ex}}^{(t)}|\mathbf{X}^{(t)},\mathbf{X}_{\sf{ex}}^{(1:t-1)}\}=p_{\mathcal{S}}(\mathbf{X}^{(t)}_{\sf{ex}}|\mathbf{X}^{(t)},\mathbf{X}^{(t-1)}_{\sf{ex}}),\label{eq:psam}\\
    &\Pr\{\mathbf{X}^{(t)}|\mathbf{X}_{\sf{ex}}^{(1:t-1)}\}\nonumber\\
    &=p_{\mathcal{L}}(\mathcal{T}_{\sf{BPE}}(\mathbf{X}^{(t)})|\mathcal{T}_{\sf{BPE}}(\mathbf{X}_{\sf{ex}}^{(t-1)}),\mathcal{T}_{\sf{BPE}}(\mathbf{r}^{(t-1)}_{\sf{ex}}),\mathbf{z}_{\sf{fix}}),\label{eq:pL}
\end{align}
\end{subequations}
where $p_{\mathcal{S}}(\cdot|\cdot)$ stands for the transition probability of the sampling operator $\mathcal{S}(\cdot)$ in \eqref{eq:sampler} and $p_{\mathcal{L}}(\cdot|\cdot)$ equals the pretrained token distribution of the LLM given in \eqref{eq:zout}. With \eqref{eq:p_ind}, we readily obtain \eqref{eq:pXex}, leading to the Markov property in \eqref{eq:markov}.

We first consider the distribution $\Pr\{\mathbf{X}_{\sf{ex}}^{(t)}|\mathbf{X}^{(t)},\mathbf{X}_{\sf{ex}}^{(1:t-1)}\}$ in \eqref{eq:psam} which describes the probability of generating the in-context examples $\mathbf{X}_{\sf{ex}}^{(t)}$. In the LLMO framework, such an operation is governed by the sampling operator $\mathcal{S}(\cdot)$. At the $(t+1)$-th iteration of Algorithm \ref{alg:alg1}, $\mathbf{X}_{\sf{ex}}^{(t)}$ is drawn from the memory $\mathbb{M}^{(t)}$. As shown in \eqref{eq:memory}, it consists of previous in-context examples $\mathbf{X}^{(t-1)}_{\sf{ex}}$ and decisions of the LLM $\mathbf{X}^{(t)}$ along with their reward values $\mathbf{r}_{\sf{ex}}^{(t-1)}$ and $\mathbf{r}^{(t)}$. Defining $\mathbf{r}_{\sf{ex}}^{(1:t-1)}\triangleq\{\mathbf{r}_{\sf{ex}}^{(\tau)}:\forall \tau=0,1,\cdots,t-1\}$ as the set of past reward vectors, it follows
\begin{subequations}\label{eq:pS}
\begin{align}
    &~~~\Pr\{\mathbf{X}_{\sf{ex}}^{(t)}|\mathbf{X}^{(t)},\mathbf{X}_{\sf{ex}}^{(1:t-1)}\}\\
    &=\Pr\{\mathbf{X}_{\sf{ex}}^{(t)}|\mathbf{X}^{(t)},\mathbf{r}^{(t)},\mathbf{X}_{\sf{ex}}^{(1:t-1)},\mathbf{r}_{\sf{ex}}^{(1:t-1)}\}\label{eq:pSb}\\
    &=p_{\mathcal{S}}(\mathbf{X}_{\sf{ex}}^{(t)}|\mathbb{M}^{(t)})\label{eq:pSd}\\
    &=p_{\mathcal{S}}(\mathbf{X}_{\sf{ex}}^{(t)}|\mathbf{X}^{(t)},\mathbf{X}_{\sf{ex}}^{(t-1)}),\label{eq:pSe}
\end{align}
\end{subequations}
where \eqref{eq:pSb} and \eqref{eq:pSe} hold since the reward values $\mathbf{r}^{(t)}$ and $\mathbf{r}_{\sf{ex}}^{(1:t-1)}$ are deterministic functions of $\mathbf{X}^{(t)}$ and $\mathbf{X}_{\sf{ex}}^{(1:t-1)}$, respectively, and \eqref{eq:pSd} results from the fact that $\mathcal{S}(\cdot)$ depends only $\mathbb{M}^{(t)}$ in \eqref{eq:memory}. We thus have \eqref{eq:psam}.

Meanwhile, the conditional distribution $\Pr\{\mathbf{X}^{(t)}|\mathbf{X}_{\sf{ex}}^{(1:t-1)}\}$ in \eqref{eq:pL} defines the probability of producing the new action population $\mathbf{X}^{(t)}$ via the pretrained LLM $p_{\mathcal{L}}(\cdot|\cdot)$ in \eqref{eq:zout}. Let $\mathbf{z}^{(t)}\triangleq\mathcal{T}_{\sf{BPE}}({\sf{pmpt}}^{(t)})$ be a token vector for the prompt $\sf{pmpt}^{(t)}$ in \eqref{eq:prompt}. It can be represented as a concatenation of three components, i.e., variable token vectors $\mathcal{T}_{\sf{BPE}}(\mathbf{X}^{(t)}_{\sf{ex}})$ and $\mathcal{T}_{\sf{BPE}}(\mathbf{r}^{(t)}_{\sf{ex}})$ and fixed token vector $\mathbf{z}_{\sf{fix}}$. Thus, $\mathbf{z}^{(t)}$ is written by
\begin{align}
    \mathbf{z}^{(t)}=[\mathcal{T}_{\sf{BPE}}(\mathbf{X}^{(t)}_{\sf{ex}}),\mathcal{T}_{\sf{BPE}}(\mathbf{r}^{(t)}_{\sf{ex}}),\mathbf{z}_{\sf{fix}}].
\end{align}
The LLM samples an output token $\mathcal{T}_{\sf{BPE}}(\mathbf{X}^{(t)})$ from $p_{\mathcal{L}}(\cdot|\mathbf{z}^{(t-1)})$. Since $\mathbf{z}^{(t-1)}$ relies only on $\mathbf{X}_{\sf{ex}}^{(t-1)}$ but not past values $\mathbf{X}_{\sf{ex}}^{(1:t-2)}$, $\mathbf{X}^{(t)}$ becomes independent of $\mathbf{X}_{\sf{ex}}^{(1:t-2)}$. Thus, we can calculate $\Pr\{\mathbf{X}^{(t)}|\mathbf{X}_{\sf{ex}}^{(1:t-1)}\}$ as
\begin{subequations}\label{eq:pLL}
\begin{align}
    &~~~\Pr\{\mathbf{X}^{(t)}|\mathbf{X}_{\sf{ex}}^{(1:t-1)}\}=\Pr\{\mathbf{X}^{(t)}|\mathbf{X}_{\sf{ex}}^{(t-1)}\}\\
    &=\Pr\{\mathcal{T}_{\sf{BPE}}(\mathbf{X}^{(t)})|\mathcal{T}_{\sf{BPE}}(\mathbf{X}_{\sf{ex}}^{(t-1)})\}\label{eq:pLLb}\\
    &=\Pr\{\mathcal{T}_{\sf{BPE}}(\mathbf{X}^{(t)})|\mathcal{T}_{\sf{BPE}}(\mathbf{X}_{\sf{ex}}^{(t-1)}),\mathcal{T}_{\sf{BPE}}(\mathbf{r}_{\sf{ex}}^{(t-1)}),\mathbf{z}_{\sf{fix}}\}\label{eq:pLLc}\\
    &=p_{\mathcal{L}}(\mathcal{T}_{\sf{BPE}}(\mathbf{X}^{(t)})|\mathbf{z}^{(t-1)}),\label{eq:pBPE}
\end{align}
\end{subequations}
where \eqref{eq:pLLb} is attained since $\mathcal{T}_{\sf{BPE}}(\cdot)$ forms an one-to-one mapping and \eqref{eq:pLLc} comes from the facts that $\mathbf{r}_{\sf{ex}}^{(t-1)}$ is directly obtained from $\mathbf{X}_{\sf{ex}}^{(t-1)}$ and $\mathbf{z}_{\sf{fix}}$ is a constant. This proves \eqref{eq:pL}.

Combining \eqref{eq:pS} and \eqref{eq:pLL}, the conditional probability $\Pr\{\mathbf{X}_{\sf{ex}}^{(t)}|\mathbf{X}_{\sf{ex}}^{(1:t-1)}\}$ can be derived as
\begin{subequations}
\begin{align}
    &\Pr\{\mathbf{X}_{\sf{ex}}^{(t)}|\mathbf{X}_{\sf{ex}}^{(1:t-1)}\}\\
    &=\!\!\!\!\!\!\!\!\!\!\sum_{\mathcal{T}_{\sf{BPE}}(\mathbf{X}^{(t)})\in\mathbb{S}}\!\!\!\!\!\!\!\!p_{\mathcal{S}}(\mathbf{X}^{(t)}_{\sf{ex}}|\mathbf{X}^{(t)},\mathbf{X}^{(t-1)}_{\sf{ex}})p_{\mathcal{L}}(\mathcal{T}_{\sf{BPE}}(\mathbf{X}^{(t)})|\mathbf{z}^{(t-1)})\\
    &=\Pr\{\mathbf{X}_{\sf{ex}}^{(t)}|\mathbf{X}_{\sf{ex}}^{(t-1)}\}.
\end{align}
\end{subequations}
This completes the proof.

\section{Proof of Lemma \ref{prop:prop4}} \label{app:appD}
We first prove the case of the elitist sampler. For $s\prec \tilde{s}$, the transition probability $p_{s\tilde{s}}$ in \eqref{eq:pij} becomes $0$ since the elitist sampler always selects the $P$ best actions and prevents the transitions from $\tilde{s}$ to $s$ such that $s\prec \tilde{s}$. This means that $\mathbf{P}_{1}$ and $\mathbf{P}_{4}$ are given as upper triangular matrices. Also, the matrix $\mathbf{P}_{3}$, which represents the transition probability from non-optimal states to optimal ones, is given by $\mathbf{P}_{3}=\mathbf{0}$. In contrast, due to the fact that $\mathbf{P}_{4}$ is upper triangular, for some $s\succ\tilde{s}$, $p_{s\tilde{s}}$ has at least one positive value, implying that $\mathbf{P}_{2}$ has at least one positive entry in each column.

Next, we discuss the LLMO with the LIFO sampling which simply chooses the in-context example as the LLM output in the previous iteration as $\mathbf{X}^{(t)}_{\sf{ex}}=\mathbf{X}^{(t-1)}$. In this case, the transition probability $p_{s\tilde{s}}$ only depends on $p_{\mathcal{L}}(\cdot|\cdot)$ of the LLM. From \eqref{eq:topk}, the probability of generating a certain state $s^{(t)}\in\mathbb{S}$ is generally positive. As a result, all submatrices have positive elements. This completes the proof.

\section{Proof of Theorem \ref{prop:prop5}}\label{app:appE}

Let us denote $\boldsymbol{\pi}^{(t)}\triangleq\{\Pr\{s^{(t)}=s\}:\forall s\in\mathbb{S}\}\in\mathbb{R}^{|\mathbb{S}|}$ as the probability vector of all states $s\in\mathbb{S}$ at the $t$-th iteration of Algorithm \ref{alg:alg1}. Also, we define $\boldsymbol{\pi}^{(t)}_{\mathbb{S}^{\star}}\triangleq\{\Pr\{s^{(t)}=s\}:\forall s\in\mathbb{S}^{\star}\}\in\mathbb{R}^{|\mathbb{S}^{\star}|}$ and $\boldsymbol{\pi}^{(t)}_{\mathbb{S}^{\prime}}\triangleq\{\Pr\{s^{(t)}=s\}:\forall s\in\mathbb{S}^{\prime}\}\in\mathbb{R}^{|\mathbb{S}^{\prime}|}$ as the probabilities of the optimal states $\mathbb{S}^{\star}$ and non-optimal states $\mathbb{S}^{\prime}$, respectively. Then, $\boldsymbol{\pi}^{(t)}$ can be written as
\begin{align}
    \boldsymbol{\pi}^{(t)}=[(\boldsymbol{\pi}^{(t)}_{\mathbb{S}^{\star}})^{T}, (\boldsymbol{\pi}^{(t)}_{\mathbb{S}^{\prime}})^{T}]^{T}.
\end{align}
According to the Markov chain, for any initial distribution $\boldsymbol{\pi}^{(0)}$, the probability vector $\boldsymbol{\pi}^{(t)}$ is obtained as
\begin{align}
    \boldsymbol{\pi}^{(t)}=\mathbf{P}_{\sf{LLM}}^{t}\boldsymbol{\pi}^{(0)}.\label{eq:pit}
\end{align}
Therefore, $\lim_{t\rightarrow\infty}\Pr\{s^{(t)}\in\mathbb{S}^{\star}\}$ can be computed as
\begin{align}\label{eq:pconv}
    \lim_{t\rightarrow\infty}\Pr\{s^{(t)}\in\mathbb{S}^{\star}\}=\sum_{s\in\mathbb{S}^{\star}}[\boldsymbol{\pi}_{\mathbb{S}^{\star}}^{(\infty)}]_{s}=1-\sum_{s\in\mathbb{S}^{\prime}}[\boldsymbol{\pi}_{\mathbb{S}^{\prime}}^{(\infty)}]_{s},
\end{align}
where $[\mathbf{u}]_{v}$ stands for the $v$-th element of a vector $\mathbf{u}$. The convergence of the LLMO is guaranteed if the probability vector of the non-optimal states becomes $\boldsymbol{\pi}_{\mathbb{S}^{\prime}}^{(\infty)}=\mathbf{0}_{|\mathbb{S}^{\prime}|}$. Otherwise, the LLMO might not identify the global optimal solution when $\boldsymbol{\pi}_{\mathbb{S}^{\prime}}^{(\infty)}$ has at least one positive element.

We first establish the optimality under the elitist sampler. Using \eqref{eq:pit}, it follows
\begin{align}
    \boldsymbol{\pi}_{\mathbb{S}^{\prime}}^{(\infty)}=\mathbf{E}\boldsymbol{\pi}^{(\infty)}=\mathbf{E}\mathbf{P}_{\sf{LLM}}^{\infty}\boldsymbol{\pi}^{(0)},\label{eq:piprime}
\end{align}
where $\mathbf{E}\triangleq[\mathbf{0}_{|\mathbb{S}^{\prime}|\times|\mathbb{S}^{\star}|},\mathbf{I}_{|\mathbb{S}^{\prime}|}]\in\mathbb{R}^{|\mathbb{S}^{\prime}|\times|\mathbb{S}|}$. With the elitist sampling, $\mathbf{P}^{\infty}_{\sf{LLM}}$ with $\mathbf{P}_{3}=\mathbf{0}$ can be calculated as \cite{MC}
\begin{align}
\mathbf{P}^{\infty}_{\sf{LLM}}=\begin{bmatrix}
        \mathbf{P}_{1}^{\infty} & \mathbf{P}_{1}^{\infty}\mathbf{P}_{2}(\mathbf{I}-\mathbf{P}_{4})^{-1} \\
        \mathbf{0}_{|\mathbb{S}^{\prime}|\times|\mathbb{S}^{\star}|} & \mathbf{P}_{4}^{\infty}
    \end{bmatrix}.
\end{align}
Recall that $\mathbf{P}_{\sf{LLM}}$ is a Markov matrix, where each column sums to 1. From Lemma \ref{prop:prop4}, we can see that with the elitist operator, $\mathbf{P}_{2}$ has at least one positive element in each column. Hence, the sum of each column in $\mathbf{P}_{4}$ is always smaller than 1. According to the Perron-Frobenius theorem \cite{PF}, the maximum eigenvalue of $\mathbf{P}_{4}$ is less than 1, implying that $\mathbf{P}_{4}^{\infty}=\mathbf{0}$. Consequently, $\boldsymbol{\pi}_{\mathbb{S}^{\prime}}^{(\infty)}$ in \eqref{eq:piprime} becomes $\boldsymbol{\pi}_{\mathbb{S}^{\prime}}^{(\infty)}=\mathbf{0}_{|\mathbb{S}^{\prime}|}$, which proves the optimality under the elitist sampler.

We now analyze the LIFO sampler. It is inferred from Lemma \ref{prop:prop4} that with the LIFO sampling, all elements of the transition matrix $\mathbf{P}_{\sf{LLM}}$ are positive. For such a positive matrix, the Perron-Frobenius theorem states that there exists a unique steady-state probability vector $\boldsymbol{\pi}^{(\infty)}\succ\mathbf{0}$ with positive entries. This indicates that $\mathbf{0}_{|\mathbb{S}^{\prime}|}\prec\boldsymbol{\pi}_{\mathbb{S}^{\prime}}^{(\infty)}\prec\mathbf{1}_{|\mathbb{S}^{\prime}|}$, which results in $\lim_{t\rightarrow\infty}\Pr\{s^{(t)}\in\mathbb{S}^{\star}\}<1$. Therefore, the LIFO sampling cannot guarantee the optimality. This completes the proof.

\end{appendices}
\bibliography{arXiv}

\begin{thebibliography}{10}

\bibitem{NLee:20}
J.~Choi, N.~Lee, S.-N. Hong, and G.~Caire, ``Joint user selection, power allocation, and precoding design with imperfect {CSIT} for multi-cell {MU-MIMO} downlink systems,'' {\em IEEE Trans. Wireless Commun.}, vol.~19, pp.~162--176, Jan. 2020.

\bibitem{WZhou:24}
W.~Zhou, D.~Zhang, M.~Debbah, and I.~Lee, ``Robust precoding designs for multiuser {MIMO} systems with limited feedback,'' {\em IEEE Trans. Wireless Commun.}, vol.~23, pp.~9583--9595, Aug. 2024.

\bibitem{BEmil:15}
E.~Björnson, L.~Sanguinetti, J.~Hoydis, and M.~Debbah, ``Optimal design of energy-efficient multi-user {MIMO} systems: {Is} massive {MIMO} the answer?,'' {\em IEEE Trans. Wireless Commun.}, vol.~14, pp.~3059--3075, Jun. 2015.

\bibitem{BBO}
P.~M. Pardalos, V.~Rasskazova, and M.~N. Vrahatis, {\em Black Box Optimization, Machine Learning, and No-Free Lunch Theorems}.
\newblock Springer, 2021.

\bibitem{Boyd}
S.~Boyd and L.~Vandenberghe, {\em {Convex Optimization}}.
\newblock Cambridge University Press, 2004.

\bibitem{BSUM}
M.~Hong, M.~Razaviyayn, Z.-Q. Luo, and J.-S. Pang, ``A unified algorithmic framework for block-structured optimization involving big data: {W}ith applications in machine learning and signal processing,'' {\em IEEE Signal Process. Mag.}, vol.~33, pp.~57--77, Jan. 2016.

\bibitem{HLee:19JSAC}
H.~Lee, S.~H. Lee, and T.~Q.~S. Quek, ``{Deep learning for distributed optimization: Applications to wireless resource management},'' {\em IEEE J. Sel. Areas Commun.}, vol.~37, pp.~2251--2266, Oct. 2019.

\bibitem{HLee:22}
H.~Lee, S.~H. Lee, and T.~Q.~S. Quek, ``Artificial intelligence meets autonomy in wireless networks: {A} distributed learning approach,'' {\em IEEE Netw.}, vol.~36, pp.~100--107, Nov. 2022.

\bibitem{HLee:22-IoT}
H.~Lee, S.~H. Lee, and T.~Q.~S. Quek, ``{MOSAIC}: {M}ultiobjective optimization strategy for {AI}-aided internet of things communications,'' {\em IEEE Internet Things J.}, vol.~9, pp.~15657--15673, Sep. 2022.

\bibitem{HLee:21}
H.~Lee, J.~Kim, and S.-H. Park, ``Learning optimal fronthauling and decentralized edge computation in fog radio access networks,'' {\em IEEE Trans. Wireless Commun.}, vol.~20, pp.~5599--5612, Sep. 2021.

\bibitem{HLee:21-TWC}
H.~Lee, S.~H. Lee, and T.~Q.~S. Quek, ``Learning autonomy in management of wireless random networks,'' {\em IEEE Trans. Wireless Commun.}, vol.~20, pp.~8039--8053, Dec. 2021.

\bibitem{GA}
J.~H. Holland, {\em Adaptation in Natural and Artificial Systems}.
\newblock Ann Arbor: The University of Michigan Press, 1975.

\bibitem{BO}
B.~Shahriari, K.~Swersky, Z.~Wang, R.~P. Adams, and N.~de~Freitas, ``Taking the human out of the loop: A review of bayesian optimization,'' {\em Proc. IEEE}, vol.~104, pp.~148--175, Jan. 2016.

\bibitem{RL}
R.~S. Sutton and A.~G. Barto, {\em {Reinforcement Learning: An Introduction}}.
\newblock MIT Press, 2nd~ed., 2018.

\bibitem{GPT2}
A.~Radford, J.~Wu, R.~Child, D.~Luan, D.~Amodei, and I.~Sutskever, ``Language models are unsupervised multitask learners,'' 2019.
\newblock [Online] Available: \url{https://cdn.openai.com/better-language-models/language_models_are_unsupervised_multitask_learners.pdf}.

\bibitem{GPT3}
T.~B. Brown, B.~Mann, N.~Ryder, M.~Subbiah, J.~D. Kaplan, P.~Dhariwal, A.~Neelakantan, {\em et~al.}, ``Language models are few-shot learners,'' Jul. 2020.
\newblock [Online] Available: https://arxiv.org/abs/2005.14165.

\bibitem{LLaMA}
H.~Touvron {\em et~al.}, ``{LLaMA}: {O}pen and efficient foundation language models,'' Feb. 2023.
\newblock [Online] Available: https://arxiv.org/abs/2302.13971.

\bibitem{OPRO}
C.~Yang, X.~Wang, Y.~Lu, H.~Liu, Q.~V. Le, D.~Zhou, and X.~Chen, ``{Large language models as optimizers},'' {\em in Proc. Int. Conf. Learn. Represent. (ICLR)}, Apr. 2024.

\bibitem{BHuang:24}
B.~Huang, X.~Wu, Y.~Zhou, J.~Wu, L.~Feng, R.~Cheng, and K.~C. Tan, ``Exploring the true potential: {E}valuating the black-box optimization capability of large language models,'' Jul. 2024.
\newblock [Online] Available: https://arxiv.org/abs/2404.06290.

\bibitem{LMEA}
S.~Liu, C.~Chen, X.~Qu, K.~Tang, and Y.-S. Ong, ``{Large language models as evolutionary optimizers},'' 2024.
\newblock [Online] Available: https://arxiv.org/abs/2310.19046.

\bibitem{GD}
P.-F. Guo, Y.-H. Chen, Y.-D. Tsai, and S.-D. Lin, ``{Towards optimizing with large language model},'' 2023.
\newblock [Online] Available: https://arxiv.org/abs/2310.05204.

\bibitem{ANie:23}
A.~Nie, C.-A. Cheng, A.~Kolobov, and A.~Swaminathan, ``Importance of directional feedback for {LLM}-based optimizers,'' in {\em Proc. Adv. Neural Inf. Process. Syst. (NeurIPS)}, pp.~1--12, Dec. 2023.

\bibitem{LEO}
S.~Brahmachary {\em et~al.}, ``{Large language model-based evolutionary optimizer: Reasoning with elitism},'' {\em Neurocomput.}, vol.~622, p.~129272, 2025.

\bibitem{MOEAD}
F.~Liu, X.~Lin, Z.~Wang, S.~Yao, X.~Tong, M.~Yuan, and Q.~Zhang, ``{Large language model for multi-objective evolutionary optimization},'' Mar. 2024.
\newblock [Online] Available: https://arxiv.org/abs/2310.12541.

\bibitem{HLee:24a}
H.~Lee, M.~Kim, S.~Baek, N.~Lee, M.~Debbah, and I.~Lee, ``{Large language models for knowledge-free network management: Feasibility study and opportunities},'' Sep. 2024.
\newblock [Online] Available: https://arxiv.org/abs/2410.17259.

\bibitem{HLee:24b}
H.~Lee, M.~Kim, S.~Baek, W.~Zhu, M.~Debbah, and I.~Lee, ``{AI-driven decentralized network management: Leveraging multi-agent large language models for scalable optimization},'' {\em IEEE Commun. Mag.}, vol.~63, pp.~50--56, Jun. 2025.

\bibitem{MALLM}
H.~Zou, Q.~Zhao, L.~Baria, M.~Bennis, and M.~Debbah, ``{Wireless multi-agent generative AI: From connected intelligence to collective intelligence},'' {\em {submitted to} IEEE Commun. Mag.}, 2023.
\newblock [Online] Available: https://arxiv.org/abs/2307.02757.

\bibitem{MALLM2}
H.~Zou, Q.~Zhao, L.~Baria, Y.~Tian, M.~Bennis, S.~Lasaulce, M.~Debbah, and F.~Bader, ``{GenAINet: Enabling wireless collective intelligence via knowledge transfer and reasoning},'' {\em submitted to IEEE Commun. Mag.}, 2024.
\newblock [Online] Available: https://arxiv.org/abs/2402.16631.

\bibitem{HLi:24}
H.~Li, M.~Xiao, K.~Wang, D.~I. Kim, and M.~Debbah, ``{Large language model based multi-objective optimization for integrated sensing and communications in UAV networks},'' Oct. 2024.
\newblock [Online] Available: https://arxiv.org/abs/2410.05062.

\bibitem{KQiu:24}
K.~Qiu, S.~Bakirtzis, I.~Wassell, H.~Song, J.~Zhang, and K.~Wang, ``Large language model-based wireless network design,'' {\em IEEE Wireless Commun. Lett.}, vol.~13, pp.~3340--3344, Dec. 2024.

\bibitem{JTong:24}
J.~Tong, J.~Shao, Q.~Wu, W.~Guo, Z.~Li, Z.~Lin, and J.~Zhang, ``Wireless{A}gent: {L}arge language model agents for intelligent wireless networks,'' Sep. 2024.
\newblock [Online] Available: https://arxiv.org/abs/2409.07964.

\bibitem{CoT}
J.~Wei, X.~Wang, D.~Schuurmans, M.~Bosma, B.~Ichter, F.~Xia, E.~H. Chi, Q.~V. Le, and D.~Zhou, ``{Chain-of-thought prompting elicits reasoning in large language models},'' {\em in Proc. Adv. Neural Inf. Process. Syst. (NeurIPS)}, pp.~24824--24837, 2022.

\bibitem{CoT-SC}
X.~Wang, J.~Wei, D.~Schuurmans, Q.~Le, E.~H. Chi, S.~Narang, A.~Chowdhery, and D.~Zhou, ``{Self-consistency improves chain of thought reasoning in language models},'' {\em in Proc. Int. Conf. Learn. Represent. (ICLR)}, Apr. 2023.

\bibitem{mirchandani2023large}
S.~Mirchandani, F.~Xia, P.~Florence, B.~Ichter, D.~Driess, M.~G. Arenas, K.~Rao, D.~Sadigh, and A.~Zeng, ``Large language models as general pattern machines,'' in {\em Proc. Annu. Conf. Robot Learn. (CoRL)}, pp.~1--21, Nov. 2023.

\bibitem{LLMSense}
X.~Ouyang and M.~Srivastava, ``{LLMS}ense: {H}arnessing {LLM}s for high-level reasoning over spatiotemporal sensor traces,'' Mar. 2024.
\newblock [Online] Available: https://arxiv.org/abs/2403.19857.

\bibitem{ReACT}
S.~Yao, D.~Yu, J.~Zhao, I.~Shafran, T.~L. Griffiths, Y.~Cao, and K.~Narasimhan, ``{R}e{A}ct: Synergizing reasoning and acting in language models,'' in {\em Proc. Adv. Neural Inf. Process. Syst. (NeurIPS)}, pp.~1--19, Dec. 2022.

\bibitem{Reflextion}
N.~Shinn, F.~Cassano, A.~Gopinath, K.~Narasimhan, and S.~Yao, ``Reflextion: {L}anguage agents with verbal reinforcement learning,'' in {\em Proc. Adv. Neural Inf. Process. Syst. (NeurIPS)}, pp.~1--19, Dec. 2023.

\bibitem{OPTIMUS}
A.~AhmadiTeshnizi, W.~Gao, and M.~Udell, ``Opti{MUS}: {S}calable optimization modeling with {(MI)LP} solvers and large language models,'' in {\em Proc. Int. Conf. Machine Learn. (ICML)}, pp.~577 -- 596, Jul. 2024.

\bibitem{KShen:22}
K.~Shen, S.~Safapourhajari, T.~De~Pessemier, L.~Martens, W.~Joseph, and Y.~Miao, ``Optimizing the focusing performance of non-ideal cell-free m{MIMO} using genetic algorithm for indoor scenario,'' {\em IEEE Trans. Wireless Commun.}, vol.~21, pp.~8832--8845, Oct. 2022.

\bibitem{Yang:24}
T.~Yang, H.~Yin, R.~Song, and L.~Zhang, ``A block quantum genetic interference mitigation algorithm for dynamic metasurface antennas and field trials,'' {\em IEEE Wireless Commun. Lett.}, vol.~13, pp.~3678--3682, Dec. 2024.

\bibitem{DDPG}
S.~Hwang, H.~Lee, J.~Park, and I.~Lee, ``{Decentralized computation offloading with cooperative UAVs: A multi-agent deep reinforcement learning perspective},'' {\em IEEE Wireless Commun.}, vol.~29, pp.~24--31, Aug. 2022.

\bibitem{MKim:24}
M.~Kim, H.~Lee, S.~Hwang, M.~Debbah, and I.~Lee, ``Cooperative multi-agent deep reinforcement learning methods for {UAV}-aided mobile edge computing networks,'' {\em IEEE Internet Things J.}, vol.~11, pp.~38040--38053, Dec. 2024.

\bibitem{ICL}
S.~M. Xie, A.~Raghunathan, P.~Liang, and T.~Ma, ``{An explanation of in-context learning as implicit Bayesian inference},'' {\em in Proc. Int. Conf. Learn. Represent. (ICLR)}, Apr. 2022.

\bibitem{ZCoT}
T.~Kojima, S.~S. Gu, M.~Reid, Y.~Matsuo, and Y.~Iwasawa, ``Large language models are zero-shot reasoners,'' in {\em Proc. Adv. Neural Inf. Process. Syst. (NeurIPS)}, pp.~22199--22213, Dec. 2022.

\bibitem{Hallucination}
Z.~Ji, N.~Lee, R.~Frieske, T.~Yu, D.~Su, Y.~Xu, E.~Ishii, Y.~J. Bang, A.~Madotto, and P.~Fung, ``{Survey of hallucination in natural language generation},'' {\em ACM Comput. Surv.}, vol.~55, pp.~1--38, Mar. 2023.

\bibitem{Hallucination2}
Z.~Ji, T.~Yu, Y.~Xu, N.~Lee, E.~Ishii, and P.~Fung, ``{Towards mitigating hallucination in large language models via self-reflection},'' {\em in Proc. Conf. Empirical Methods Natural Lang. Process. (EMNLP)}, 2023.

\bibitem{GRudolph:94}
G.~Rudolph, ``Convergence analysis of canonical genetic algorithms,'' {\em IEEE Trans. Neural Netw.}, vol.~5, pp.~96--101, Jan. 1994.

\bibitem{JSuzuki:95}
J.~Suzuki, ``A {M}arkov chain analysis on simple genetic algorithms,'' {\em IEEE Trans. Syst. Man, Cybern.}, vol.~25, pp.~655--659, Apr. 1995.

\bibitem{HJun:16}
J.~He and G.~Lin, ``Average convergence rate of evolutionary algorithms,'' {\em IEEE Trans. Evol. Comput.}, vol.~20, pp.~316--321, Apr. 2016.

\bibitem{ACR0}
R.~S. Varga, {\em {Matrix Iterative Analysis}}.
\newblock Springer, 2009.

\bibitem{ACR1}
J.~He and L.~Kang, ``On the convergence rates of genetic algorithms,'' {\em Theore. Comput. Sci.}, vol.~229, pp.~23--39, Nov. 1999.

\bibitem{ACR2}
Y.~Chen and J.~He, ``Average convergence rate of evolutionary algorithms in continuous optimization,'' {\em Inf. Sci.}, vol.~562, pp.~200--219, Jul. 2021.

\bibitem{WMMSE}
Q.~Shi, M.~Razaviyayn, Z.-Q. Luo, and C.~He, ``An iteratively weighted {MMSE} approach to distributed sum-utility maximization for a {MIMO} interfering broadcast channel,'' {\em IEEE Trans. Signal Process.}, vol.~59, pp.~4331--4340, Sep. 2011.

\bibitem{GNN}
H.~Lee, S.~H. Lee, and T.~Q.~S. Quek, ``Learning autonomy in management of wireless random networks,'' {\em IEEE Trans. Wireless Commun.}, vol.~20, pp.~8039--8053, Dec. 2021.

\bibitem{FP}
K.~Shen and W.~Yu, ``Fractional programming for communication systems—{P}art {I}: {P}ower control and beamforming,'' {\em IEEE Trans. Signal Process.}, vol.~66, pp.~2616--2630, May 2018.

\bibitem{PF}
R.~Horn and C.~Johnson, {\em {Matrix Analysis}}.
\newblock Cambridge University Press, 1985.

\bibitem{Vladik:93}
V.~Kreinovich, C.~Quintana, and O.~Fuentes, ``{G}enetic algorithms: {W}hat fitness scaling is optimal?,'' {\em Cybern. Syst.}, vol.~24, no.~1, pp.~9--26, 1993.

\bibitem{DeepMIMO}
A.~Alkhateeb, ``{DeepMIMO}: A generic deep learning dataset for millimeter wave and massive {MIMO} applications,'' in {\em Proc. Inf. Theory Appl. Workshop (ITA)}, pp.~1--8, Feb. 2019.

\bibitem{TelecomGPT}
H.~Zou, Q.~Zhao, Y.~Tian, L.~Bariah, F.~Bader, T.~Lestable, and M.~Debbah, ``Telecom{GPT}: {A} framework to build telecom-specfic large language models,'' Jul. 2024.
\newblock [Online] Available: https://arxiv.org/abs/2407.09424.

\bibitem{MC}
M.~Isoifescu, {\em {Finite Markov Processes and Their Applications}}.
\newblock Chichester: Wiley, 1980.

\end{thebibliography}
\bibliographystyle{ieeetr}

\end{document}